\documentclass[aps, prd, 12pt, superscriptaddress, amsmath, amssymb, tightenlines]{revtex4-2}
\pdfoutput=1
\usepackage{amsmath}
\usepackage{amssymb}
\usepackage[utf8]{inputenc}
\usepackage[unicode,breaklinks=true]{hyperref}
\usepackage{graphicx}
\allowdisplaybreaks[1]

\begin{document}

\title{
Higher order corrections to beyond-all-order effects in a fifth order
Korteweg-de Vries equation
}

\author{Gyula Fodor}
\affiliation{Wigner Research Centre for Physics, 1525 Budapest 114, P.O.~Box 49, Hungary}
\author{Péter Forgács}
\affiliation{Wigner Research Centre for Physics, 1525 Budapest 114, P.O.~Box 49, Hungary}
\affiliation{Institut Denis-Poisson CNRS/UMR 7013, Université de Tours, Parc de
Grandmont, 37200 Tours, France}
\author{Muneeb Mushtaq}
\affiliation{Wigner Research Centre for Physics, 1525 Budapest 114, P.O.~Box 49, Hungary}
\affiliation{Institute for Theoretical Physics, Eötvös University,
Pázmány Péter sétány 1/A, H-1117 Budapest, Hungary}

\begin{abstract}
\setlength{\baselineskip}{1.1\baselineskip}
A perturbative scheme is applied to calculate corrections to
the leading, exponentially small (beyond-all-orders) amplitude of the ``trailing'' wave asymptotics of weakly localized solitons. The model considered is a
Korteweg-de Vries equation modified by a fifth order derivative term, $\epsilon^2\partial_x^5$ with $\epsilon\ll1$ (fKdV).
The leading order corrections to the tail amplitude are calculated up to ${\cal{O}}(\epsilon^5)$. An arbitrary precision numerical code is implemented
to solve the fKdV equation and to check the perturbative results. Excellent agreement is found between the numerical and analytical results. Our work also clarifies the origin of a long-standing disagreement between the ${\cal{O}}(\epsilon^2)$ perturbative result of Grimshaw and Joshi [SIAM J. Appl. Math. 55, 124 (1995)] and the numerical results of Boyd [Comp. Phys. 9, 324 (1995)].
\end{abstract}

\maketitle

\setlength{\baselineskip}{1.2\baselineskip}

\section{Introduction}

Oscillons -- slowly radiating lumps in theories containing scalar fields with a well-defined core and living for very long times  -- have generated quite some interest in view of their numerous physical applications \cite{Copeland95,Honda2001}; for recent reviews see   \cite{Cyncynates2021,Fodor19,Visinelli21} and the references cited therein.
An important problem is to determine the radiation rate and lifetimes of oscillons; see Refs.\ \cite{Zhang2020,Olle2020} for recent results and also for further references.
In Refs.\ \cite{Fodor09a,Fodor09b,Fodor19} the perturbative scheme of Ref.\ \cite{SegurKruskal87} has been generalized to a large class of theories, yielding the leading order estimate for oscillon lifetimes. The main idea has been to approximate slowly radiating oscillons through the adiabatic evolution of appropriate \textsl{stationary} configurations, called ``quasibreathers'' -- weakly localized lumps with asymptotic standing wave tails \cite{Fodor06,Saffin07}. Quasibreathers are time-periodic and can be thought of as oscillons made stationary by incoming radiation from infinity. Determination of the amplitude of the quasibreather wave tails is necessary to deduce the radiation rate of time-dependent oscillons. A rather nontrivial aspect of the perturbative computation of the standing wave tail amplitude for quasibreathers corresponding to long-lived oscillons is that it is beyond all orders in perturbation theory.

It still remains a challenging problem to calculate higher order corrections to the oscillating tails of
quasibreathers. To prepare the ground for such higher order computations we have taken up a much simpler problem
- the computation of higher order corrections to asymptotic wave tails in the familiar Korteweg-de Vries (KdV) equation
modified by a fifth order derivative term (fKdV), also called the Kawahara equation \cite{Kakutani69,Kawahara72}.
The fKdV equation plays an important role in many applications in plasma physics and in hydrodynamics.
For a detailed derivation of the fKdV equation in a hydrodynamical context see \cite{Hunter88}. A crucial point of interest of the fKdV equation from our point of view is that the familiar solitary wave solutions of the KdV equation are deformed into oscillon-type objects, losing continuously some of their mass by radiating small amplitude waves in the direction of propagation \cite{Benilov93}.
It has been proven that the spatially localized solitary traveling wave solution of the KdV equation ceases to exist when a fifth order dispersion term proportional to $\epsilon^2$ is added \cite{Amick91,Gunney99}.
What happens is that the extra dispersion term causes the KdV solitary wave to develop a radiating tail whereby it loses energy \cite{Benilov93}.
Bounded, stationary solutions of the fKdV equation are weakly localized, in that asymptotically they tend to a standing wave ``tail''. This is to be contrasted to the exponential falloff of the well localized KdV solitons. In this context such weakly localized, stationary solutions, baptized ``nanopterons'' have been studied systematically in Refs.\ \cite{Boyd90,Boyd98}.

For both the quasibreathers and the nonlocal solitons of the fKdV equation, the tail amplitude is exponentially small in terms of a parameter ($\epsilon$) which characterizes the specific perturbation or describes the core amplitude (for quasibreathers), and to leading order it can be written as
\begin{equation}\label{leading}
 \alpha_m=\frac{\lambda}{\epsilon^{\nu}}\exp\left(-\frac{\sigma}{\epsilon}\right)\,,
\end{equation}
where $\lambda$, $\nu$ and $\sigma$ are constants. For the fKdV case the exponent in the denominator is $\nu=2$, $\sigma=\pi/2$ (in suitable units) while for spherically symmetric oscillons $\nu=(d-1)/2$ where $d$ is the number of spatial dimensions. The proportionality constant $\lambda$ is rather nontrivial to compute. Following the technique pioneered in Ref.\ \cite{SegurKruskal87}, the value of $\lambda$ can be obtained by going through a complicated asymptotic matching calculation in the complex plane. For the fKdV equation the leading order result has been obtained in Ref.\ \cite{Pomeau88} with the result $\lambda\approx19.969\pi$.

It is desirable to compute higher order corrections to the leading order result \eqref{leading} from both a theoretical and a practical point of view. One would like to establish a systematic framework for such calculations, quantify the contributions of higher order terms, but it is also necessary to know higher order corrections in order to compare the results to numerical simulations for small, but obviously finite values of $\epsilon$. For very small $\epsilon$ values where the leading order term is supposed to dominate, the tail amplitude becomes so small that it is extremely demanding and difficult to calculate by numerical methods. This emphasizes the necessity
to investigate corrections to Eq.\ \eqref{leading}. For the fKdV problem and also for quasibreathers, it is to be expected that the leading order result \eqref{leading} gets several types of corrections, with the dominant ones involving only $\epsilon$. In the simplest case such corrections can be written as a power series of $\epsilon$ of the form
\begin{equation}
 \alpha_m=\frac{\lambda}{\epsilon^{\nu}}
 \exp\left(-\frac{\sigma}{\epsilon}\right)
 \left(1+\zeta_1\epsilon+\zeta_2\epsilon^2+\zeta_3\epsilon^3
 +\cdots\right)\,, \label{eqallambmusig}
\end{equation}
where $\zeta_j$ are some yet unknown constants. (For $d\geq 2$ dimensional oscillons there are also $\epsilon\ln\epsilon$ and similar logarithmic terms \cite{Fodor09b}.) For most cases including the scalar field problem the tail amplitude has been calculated only to leading order; i.e., even $\zeta_1$ is unknown yet.
On the other hand, for the fKdV equation detailed properties of weakly nonlocal solitons
have been under intense scrutiny, and even the computation of $\zeta_1$ and $\zeta_2$ was undertaken by Grimshaw and Joshi in Ref.\ \cite{Grimshaw95}. Since we expect that similar methods to that of Ref.\ \cite{Grimshaw95} will be applicable for the more complicated scalar field problem as well, a thorough understanding of the higher order contributions in the fKdV problem appears to be necessary to us, which is the subject of the present paper.

According to the results of Ref.\ \cite{Grimshaw95}, the first two corrections in Eq.\ \eqref{eqallambmusig} are given as $\zeta_1=-\pi$ and $\zeta_2=\pi^2/2\approx 4.935$ in suitable units.
The value of $\zeta_1$ is in agreement with the numerical computations by Boyd \cite{Boyd91,Boyd95,Boyd98}, using a multiple precision pseudospectral code. According to the numerical estimates of Boyd, however,  $-0.1<\zeta_2<0$, which is clearly in disagreement with the result of Ref.\ \cite{Grimshaw95}. As far as we know, the reason for this discrepancy has remained unknown up to now.

We have succeeded to compute higher order corrections in Eq.\ \eqref{eqallambmusig} up to order $\epsilon^5$,
and while confirming the result for the value of $\zeta_1=-\pi$ of Ref.\ \cite{Grimshaw95}, we have found $\zeta_2=\pi^2/2-5\approx -0.065198$, a value which is quite consistent with the numerical estimates of Ref.\ \cite{Boyd95}.

To check our analytical results we have also developed a multiple precision pseudospectral numerical code to solve the fKdV equation. Since the tail amplitude for the intended $\epsilon$ parameter range can be many orders of magnitude smaller than $10^{-16}$, the use of arbitrary precision arithmetic is indispensable. The speed of present day personal computers and the efficiency of the available multiple precision arithmetic libraries allows us to reach significantly higher precision than what was available earlier. The running time of our code also allows us to apply a numerical minimization algorithm to find the phase of the tail where the tail amplitude is really minimal. This was not included in earlier publications.
Comparison of the coefficients of the higher order corrections, $\zeta_1\ldots\zeta_5$ with the results of our numerical simulations shows remarkably good agreement. In particular the numerical value of $\zeta_2$ agrees with our perturbative result to five digits of precision, leaving little doubt as to its correctness.
We have also obtained a very high order ($\sim 100$) asymptotic expansion in $\epsilon$ of the phase of the minimal amplitude wave tail in the Wentzel–Kramers–Brillouin (WKB) approximation. We could compare up to order 15 the expansion of the phase with the numerical results and an excellent agreement has been found.

The plan of the paper is the following : In Sec.~\ref{sec:fKdV} the fifth order KdV equation is introduced and its basic properties are discussed. Next, in Sec.~\ref{sec:num} our implementation of the spectral method is described in some detail.
In Sec.~\ref{secexpcore} the small $\epsilon$ expansion for the ``core'' part of the solution is carried out.
In Sec.~\ref{secexpcorr} in the framework of the WKB approximation, the linearized solution around the core is determined to arbitrary order.
The amplitude of the wave tail is given up to order 6, and the phase is given explicitly up to order 11.
Sections \ref{sec:match} and \ref{secfourth} contain the most important calculations, namely carrying out the asymptotic matching to fourth order in $\epsilon$ of the complex extension of the ``inner'' and of the ``outer'' part of the solution, near the first singularity in the complex plane. Section \ref{sec:conc} contains our conclusions.

\section{Fifth order KdV equation}\label{sec:fKdV}

The Korteweg-deVries equation modified by a small fifth derivative term can be written as \cite{Hunter88,Pomeau88}
\begin{equation}
 \epsilon^2u_{yyyyy}+u_{yyy}+6uu_y+u_t = 0 \,, \label{epuyyyyy}
\end{equation}
where $u$ is a function of the time $t$ and spatial coordinate $y$, and $\epsilon$ is a small non-negative parameter. The indices denote derivatives with respect to $t$ and $y$. We only consider stationary solutions traveling with speed $c$ to the right, so that $u$ is time independent when $x=y-ct$ is used as a comoving spatial coordinate,
\begin{equation}
 \epsilon^2u_{xxxxx}+u_{xxx}+(6u-c)u_x = 0 \,. \label{epuxfive}
\end{equation}
This form clearly shows that if $u$ is a solution and $u_c$ is a constant, then $u+u_c$ is also a solution moving with speed $c+6u_c$. The equation can be integrated once, yielding
\begin{equation}
 \epsilon^2u_{xxxx}+u_{xx}+3u^2-cu = M \,, \label{epuxxxxm}
\end{equation}
where $M$ is a constant that can be interpreted as the mass flux \cite{Grimshaw95}.
If $u$ and its derivatives tend to zero in either the positive or negative $x$ direction, then necessarily $M=0$. For $\epsilon>0$ bounded solutions necessarily have an asymptotic oscillating tail in both directions. In this case $M$ becomes determined by the boundary conditions. Two different boundary conditions have been used in the literature. One possibility, used by Boyd \cite{Boyd91,Boyd95,Boyd98}, is to impose no mass flux by requiring $M=0$. Another choice, applied by Grimshaw and Joshi \cite{Grimshaw95}, is to set the integral of $u$ zero for an interval of $x$ corresponding to the wavelength of one oscillation, thereby keeping the fluid volume constant with respect to the $u=0$ solution. Even in this latter case $M$ is extremely small, of order $\alpha^2$, where $\alpha$ is the tail amplitude, which turns out to be exponentially small in $\epsilon$. An $\alpha^2$ order small shift in $u$ and $c$ can be used to set $M$ zero. In the present paper we are interested in solutions for relatively small values of $\epsilon$.
From now on we set $M=0$, and our aim is to solve the equation
\begin{equation}
 \epsilon^2u_{xxxx}+u_{xx}+3u^2-cu = 0 \,, \label{equxxxx}
\end{equation}
for positive values of the two parameters $\epsilon$ and $c$.

Equation \eqref{equxxxx} remains invariant under the rescalings
\begin{equation}
 u=\xi^2\bar u \ \,, \quad
 x=\frac{1}{\xi}\bar x \ \,, \quad
 c=\xi^2\bar c \ \,, \quad
 \epsilon=\frac{1}{\xi}\bar\epsilon \ \,, \label{equxceresc}
\end{equation}
for any $\xi>0$ constant. Note that $\epsilon^2 c$ remains invariant. It would be possible to use this freedom to scale either of the constants $\epsilon$ or $c$ to some given value, and consider \eqref{equxxxx} as an equation containing only the other parameter. Following Grimshaw and Joshi \cite{Grimshaw95}, we use this rescaling freedom to set the speed parameter $c$ to a specific known function of $\epsilon$. The actual form of the function $c(\epsilon)$ will be fixed a bit later by the requirement that the spatial decay rate of the core of the solution should be parameter independent. Still, with this choice we reduce the number of parameters in the problem from two to only one, keeping only $\epsilon$.

We denote the value of $c\equiv c(\epsilon)$ in the $\epsilon\to 0$ limit by $c_0$. For $\epsilon=0$ and $c_0>0$ Eq.~\eqref{equxxxx} has the well-known KdV solitary wave solution
\begin{equation}
 u_0 = 2\gamma^2\mathrm{sech}^2(\gamma x) \,, \label{equ0twogamsq}
\end{equation}
where the positive $\epsilon$ independent constant $\gamma$ is defined by
\begin{equation}
 c_0 = 4\gamma^2 \,. \label{eqcnulgam}
\end{equation}
Localized asymptotically decaying solutions of the KdV equation only exist for $c_0>0$. Since our aim is to look for solutions that are as similar to solitary waves as possible, we assume that both $c_0$ and $c(\epsilon)$ are positive.

We are looking for solutions for which asymptotically $u$ becomes small. The linearization of Eq.~\eqref{equxxxx} around $u=0$ has four independent solutions. The exponentially decaying or growing solutions have the form $u=\exp(2\tilde\gamma x)$, where
\begin{equation}
 \epsilon^2(4\tilde\gamma^2)^2+4\tilde\gamma^2-c=0 \,. \label{eqepepgt}
\end{equation}
Since we assume $c>0$, this has two real valued solutions for $\tilde\gamma$, determined by
\begin{equation}
 4\tilde\gamma^2=\frac{-1+\sqrt{1+4\epsilon^2c}}{2\epsilon^2}
 \,. \label{eqgamtil}
\end{equation}
The remaining two oscillating solutions can be written in the form $u=\exp(ikx/\epsilon)$. This is a natural parametrization, since the spatial frequency tends to infinity when $\epsilon$ goes to zero. The linearization of \eqref{equxxxx} gives two real solutions for $k$, given by
\begin{equation}
 \frac{k^2}{\epsilon^2}=\frac{1+\sqrt{1+4\epsilon^2c}}{2\epsilon^2}
 \,. \label{eqkkpepep}
\end{equation}
Subtracting \eqref{eqgamtil} from \eqref{eqkkpepep} and rearranging, we obtain the simple relation between $k$ and $\tilde\gamma$,
\begin{equation}
 k^2=1+4\epsilon^2\tilde\gamma^2 \,. \label{eqkk2onep}
\end{equation}

Taking the $\epsilon\to 0$ limit of Eq.~\eqref{eqepepgt} and comparing with \eqref{eqcnulgam}, it follows that $\tilde\gamma$ tend to $\gamma$ when $\epsilon$ goes to zero. Grimshaw and Joshi \cite{Grimshaw95} made the choice to impose that $\tilde\gamma$ is $\epsilon$ independent, in which case obviously $\tilde\gamma=\gamma$. This is a natural option, since it makes the decay rate of the core independent of the parameter $\epsilon$, and the expansion formalism becomes considerably simpler. We will also assume $\tilde\gamma=\gamma$ from now on. As a consequence of this choice, the parameter $c$ necessarily becomes $\epsilon$ dependent, and from \eqref{eqepepgt} we obtain
\begin{equation}
 c=4\gamma^2+16\gamma^4\epsilon^2 \,. \label{eqc4gamp}
\end{equation}
This is the specific $c\equiv c(\epsilon)$ function that we ensure by the appropriate use of the rescaling in \eqref{equxceresc}.

Since we are looking for real solutions, instead of the exponential form $u=\exp(ikx/\epsilon)$ for the tail we write
\begin{equation}
 u=\alpha_\pm\sin\left(\frac{k|x|}{\epsilon}
 -\delta_\pm\right) \,. \label{equtailpm}
\end{equation}
The amplitude $\alpha_{+}$ and phase $\delta_{+}$ in the positive direction can be different from that of $\alpha_{-}$ and $\delta_{-}$ in the negative direction. Using the trigonometric form it is enough to use the positive root in \eqref{eqkk2onep}, hence in the following we will set
\begin{equation}
 k=\sqrt{1+4\gamma^2\epsilon^2} \,. \label{eqksqrt}
\end{equation}

A conserved quantity defined as
\begin{equation}
 F=-\frac{1}{2}cu^2+2u^3+uu_{xx}-\frac{1}{2}u_{x}^2+\epsilon^2\left(
 uu_{xxxx}-u_{x}u_{xxx}+\frac{1}{2}u_{xx}^2
 \right) \,,
\end{equation}
can be interpreted as an energy flux \cite{Benilov93,Grimshaw95}.
Using \eqref{epuxfive} it is easy to check that $F_x=0$. Substituting the form \eqref{equtailpm} of the tail, to leading order the energy flux turns out to be $F=\alpha_\pm^2/(2\epsilon^2)$. Since $F$ is constant, it follows that the amplitude of the tail at the two directions must necessarily agree, at least for small amplitudes. The phases can also be made to agree by a small shift in $x$, but the argument is not giving any information about the symmetry of the core. Numerical simulations also support the conjecture that solutions which have a single large core and small tails in both the positive and the negative directions are necessarily reflection symmetric.
The important consequence is that there are no solutions for which there is a small tail in one direction and exponential decay without any tail in the other direction. Our numerical simulations show that for each small $\epsilon$ there is a solution which decays exponentially to zero for $x>0$, and it has a core region similar to the KdV solitary wave, but continuing further to the negative direction the solution blows up at some finite $x<0$ before a standing wave tail could appear. Hence in the following we consider only solutions which are symmetric with respect to $x=0$. For large $x>0$ their tail is characterized by the parameters $\alpha$ and $\delta$,
\begin{equation}
 u=\alpha\sin\left(\frac{kx}{\epsilon}
 -\delta\right) \,. \label{equtail}
\end{equation}

Even after using the rescaling \eqref{equxceresc} to ensure the intended $\epsilon$ dependence of $c$ according to \eqref{eqc4gamp}, there is still a remaining freedom to set $c_0$ to any desired positive value. It is sufficient to solve Eq.~\eqref{equxxxx} numerically or analytically for only one special choice of $c_0$, since other solutions can be obtained by the above rescalings. The choice made by Boyd in \cite{Boyd95,Boyd98} is $c_0=4$, which is a natural option since it corresponds to $\gamma=1$. For easier comparison with Boyd's results we will also use $c_0=4$ in our numerical calculations.

In this paper we mostly use the notations introduced by Grimshaw and Joshi in \cite{Grimshaw95}, but we discuss the connection with the variables and equations used by Boyd \cite{Boyd91,Boyd95,Boyd98}. Applying the rescaling \eqref{equxceresc} with $\xi=1/\epsilon$ we get $\bar\epsilon=1$, and hence the $\epsilon^2$ factor disappears in front of the fourth derivative term. Introducing a further rescaled function by $\bar u=\bar v/6$, the coefficient of the quadratic term can change, and we obtain the form of the fKdV equation used by Boyd,
\begin{equation}
 \bar v_{\bar x\bar x\bar x\bar x}+\bar v_{\bar x\bar x}
 +\frac{1}{2}\bar v^2-\bar c\bar v = 0 \,. \label{equboyd}
\end{equation}
In this case the only parameter is $\bar c=\epsilon^2 c=4\gamma^2\epsilon^2+16\gamma^4\epsilon^4$.

\section{Numerical method}\label{sec:num}

We apply a pseudospectral numerical method (see, e.g., \cite{Boyd13}) to solve Eq.~\eqref{equxxxx}, looking for solutions $u$ that are symmetric at $x=0$. Since the tail has infinitely many oscillations, spatial compactification with standard Chebyshev expansion cannot be used efficiently in this case. Boyd \cite{Boyd91,Boyd95,Boyd98} used an additional basis function to represent the oscillating tail. In our numerical simulations we match the solution to the tail given in \eqref{equtail} at the outer numerical boundary $x=L$. By appropriate rescalings it is always possible to arrange that $c$ is given by \eqref{eqc4gamp} with $\gamma=1$. For any chosen parameter $\epsilon$ and phase $\delta$ the numerical problem can be solved to obtain a corresponding tail amplitude $\alpha$. We use two boundary conditions at $x=L$,
\begin{align}
 &\epsilon^2 u_{xx}+k^2 u=0 \,, \\
 &\epsilon\sin\left(\frac{kx}{\epsilon}-\delta\right)u_x
 -k\cos\left(\frac{kx}{\epsilon}-\delta\right)u=0 \,.
\end{align}
Together with the symmetry assumption these conditions make the solution unique.

We introduce an alternative independent variable $\theta$ by $x=L\cos\theta$. The center $x=0$ corresponds to $\theta=\pi/2$, and the outer boundary to $\theta=0$. We fix some order $N$, and represent the function $u$ by $N$ Fourier components $U_n$,
\begin{equation}
 u=\sum_{n=0}^{N-1}\frac{1}{\rho_n} U_n\cos(2n\theta) \,, \label{equfuncfour}
\end{equation}
where
\begin{equation}
 \rho_n=\left\{\begin{array}{rl}
  1 & \text{if} \ \ 1 \leq n \leq N-2 \,, \\
  2 & \text{if} \ \ n=0 \ \ \text{or} \ \ n=N-1 \,.
  \end{array}\right.
\end{equation}
We only include even Fourier components because of the reflection symmetry at the center. Actually, this corresponds to expansion in even indexed Chebyshev polynomials $T_{2n}(x/L)$, since $\cos(n\theta)=T_n(\cos\theta)$. The solution can be alternatively represented by its values at $N$ collocation points $x_n=L\cos\theta_n$, where $\theta_n=\pi n/(2(N-1))$ for $0\leq n\leq N-1$. For the function values we introduce the notation $\tilde u_n=u(x_n)$. The value of $u$ at the center is $\tilde u_{N-1}$, while at the outer boundary it is $\tilde u_0$. Using \eqref{equfuncfour}, the collocation values can be obtained by matrix multiplication,
\begin{equation}
 \tilde u_n=\sum_{j=0}^{N-1}C_{nj}U_j \,,\quad \qquad
 C_{nj}=\frac{1}{\rho_j}\cos\frac{nj\pi}{N-1} \,.
\end{equation}
The inverse transformation is
\begin{equation}
 U_n=\sum_{j=0}^{N-1}F_{nj}\tilde u_j \,,\quad \qquad
 F_{nj}=\frac{2}{N-1}C_{nj} \,.
\end{equation}
This corresponds to calculating the Fourier coefficients by numerical integration based on the collocation points. In both directions, the transformation corresponds to type I discrete cosine transform, DCT-I.

Multiplication of functions can easily be calculated using the collocation values, while derivatives can be naturally obtained using Fourier coefficients. The Fourier coefficients $(U_{xx})_n$ of the second derivative of the function $u$ represented by $U_n$ can be calculated as
\begin{equation}
 (U_{xx})_n=\sum_{j=0}^{N-1}D^{(2)}_{nj}U_j \,,
\end{equation}
where
\begin{equation}
 D^{(2)}_{nj}=\left\{\begin{array}{cl}
  \displaystyle\frac{8}{\rho_j L^2}j(j^2-n^2)
  & \text{if} \ \   j\geq n+1 \,, \\
  0 & \text{if} \ \   j< n+1 \,.
  \end{array}\right.
\end{equation}
The second derivative matrix in the collocation picture can be obtained as $\tilde D^{(2)}_{nj}=C_{nl}D^{(2)}_{lp}F_{pj}$, where the summation for repeated indices is understood. The fourth derivative matrix can be obtained most easily by multiplying the second derivative matrix by itself, $D^{(4)}_{nj}=D^{(2)}_{nl}D^{(2)}_{lj}$. For the boundary condition we also need the first derivative, but only at the boundary. This can be calculated by the scalar product $V_n\tilde u_n$, where \begin{equation}
 V_n=\sum_{j=0}^{N-1}\frac{4j^2}{L}F_{jn} \,.
\end{equation}

Since Eq.~\eqref{equxxxx} is nonlinear, we use an iterative procedure, called the Newton-Kantorovich method to solve it (see e.g.~Appendix C of \cite{Boyd13}). Suppose that at the $n$th step we have an approximate solution $u^{(n)}$. The next approximation shall be $u^{(n+1)}=u^{(n)}+\Delta$. Substituting into \eqref{equxxxx} and linearizing for $\Delta$ we obtain the equation
\begin{equation}
 \epsilon^2\Delta_{xxxx}+\Delta_{xx}
 +6u^{(n)}\Delta-c\Delta = R \,, \label{eqddeltalin}
\end{equation}
where the residual is
\begin{equation}
 R=-\epsilon^2 u^{(n)}_{xxxx}-u^{(n)}_{xx}-3\left(u^{(n)}\right)^2
 +cu^{(n)} \,.
\end{equation}
According to our experience, less than ten steps of iteration is enough to get extremely high precision solutions for this problem. The iteration can be started from the KdV solitary wave solution \eqref{equ0twogamsq}. We consider \eqref{eqddeltalin} as $N$ algebraic equations at the collocation points. The left-hand side can be considered as a linear matrix operator $L_{nj}$ multiplying the collocation values $\Delta_j$ of the unknown function $\Delta$. The collocation values $R_n$ of residuals can be calculated at each step from the previous approximation. We replace two lines of $L_{nj}$ by values enforcing the boundary conditions, while replacing the corresponding elements in $R_n$ by the previous error in the boundary conditions. The result turns out to be quite insensitive of which lines we replace; for example we can use line $0$ and $N-2$. After this, the matrix equation can be solved for $\Delta_j$. Updating $u^{(n)}$ by adding the calculated $\Delta$ yields the next approximation.

We have written equivalent C and C++ codes to solve the numerical problem. Since we are interested in comparing the numerical results to the analytical ones, we intend to calculate the tail amplitude $\alpha$ for relatively small $\epsilon$ values. It can be seen easily that the usual $16$ or $19$ digits arithmetic is not enough for our aims. The core amplitude is always close to $2$. If the tail amplitude $\alpha$ is of the order $10^{-a}$, and we intend to calculate it to $b$ digits of precision, then the whole numerical procedure should be carried out with at least $a+b$ digits of precision. There are freely available numerical packages for calculations with arbitrarily many digits of precision that are fast enough for our purposes. For our C++ codes we have used the Class Library for Numbers (CLN) \cite{clnweb}. Our codes using the C library for arbitrary-precision ball arithmetic (ARB) \cite{arbweb,Johansson17} turn out to about 20 times faster, due to more advanced matrix manipulation methods. However, writing programs using ARB is more difficult, since it requires separate lines of codes for each elementary algebraic manipulation, such as addition or multiplication of numbers.

\begin{figure}[!hbt]
 \centering
 \includegraphics[width=115mm]{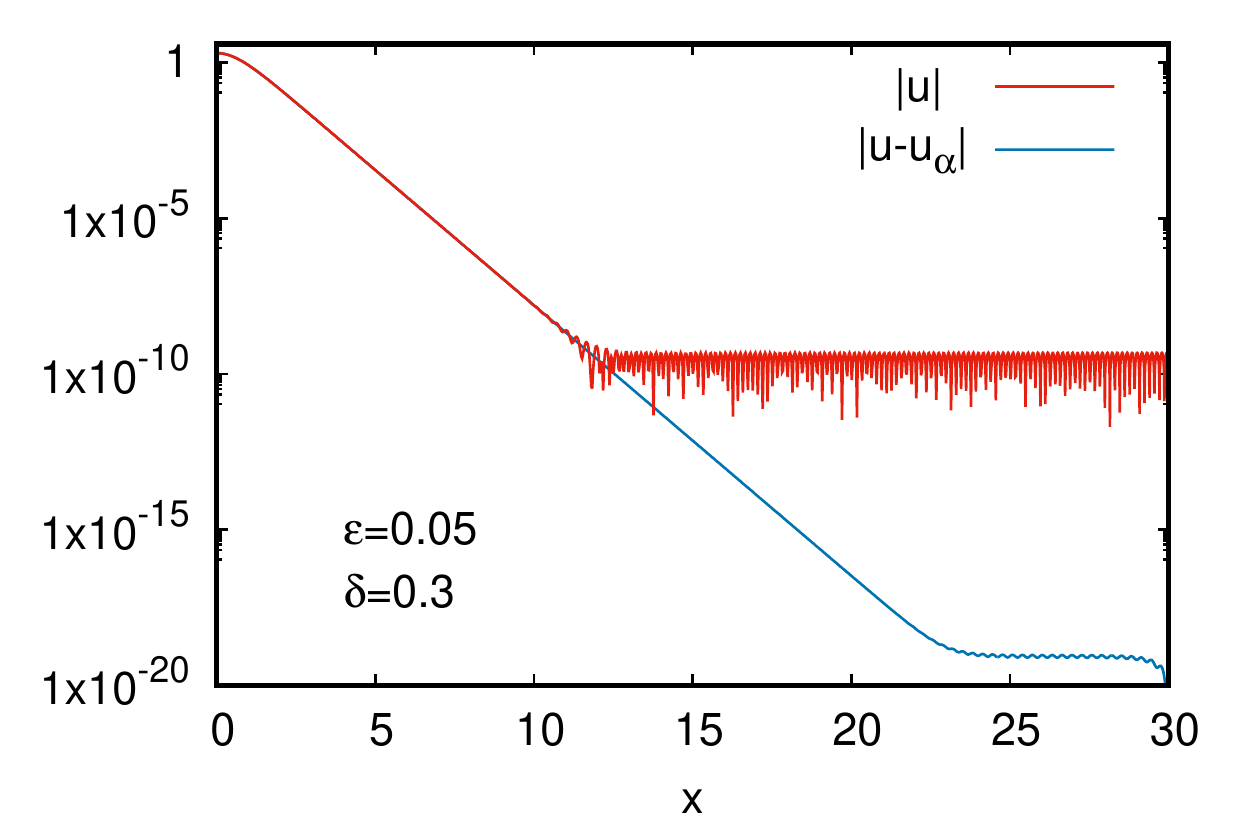}
\caption{\label{fig05log} The red curve shows $u$ as a function of $x$. Subtracting the function $u_\alpha=\alpha\sin(kx/\epsilon-\delta)$ corresponding to the matched tail we obtain the blue curve.}
\end{figure}

To illustrate the structure and precision of the obtained solutions we present some results for $\epsilon=0.05$. We use $c_0=4$ (and consequently $\gamma=1$) in all our numerical work. For the phase we choose $\delta=6\gamma\epsilon=0.3$ for this example, which corresponds to the linear approximation of the phase belonging to the minimal tail amplitude [see \eqref{eqdeltildeexp} and \eqref{eqtildel1} for details on that]. To include a large enough portion of the tail we set the outer boundary at $L=30$. In Fig.~\ref{fig05log} we plot logarithmically the $x$ dependence of the function $u$.
The downward spikes correspond to zero crossings in the tail. To illustrate the precision of the numerical solution and the matching, we also plot the difference of $u$ and the inward continuation of the matched tail \eqref{equtail}, using the tail amplitude $\alpha=4.811363414\cdot10^{-10}$ provided by the numerical code. The extremely good agreement of $u$ and the tail for $x>23$ indicates that the obtained $\alpha$ is precise to $10$ digits. To reach this precision we need at least $N=500$ collocation points and $26$ digits of precision during the whole numerical calculation. At this resolution the running time on our desktop computer is about $90$ s for the CLN code, and only $2$ s for the ARB code. Even if we increase the resolution further, the difference from the matched tail will not go below $10^{-20}$ for the present $\epsilon$ value. The reason for this is that we represent the matched tail with the linearized solution \eqref{equtail}. If the tail amplitude is $\alpha$ then we make an error of order $\alpha^2$ by this choice.

Numerical calculation of symmetric solutions with very small tails have been first reported by Boyd in \cite{Boyd91}, applying spectral methods. In that paper the solution is matched to a higher order nonlinear representation of the tail. Boyd called the method cnoidal matching, since in the $\epsilon=0$ case the spatially periodic higher amplitude KdV solutions are given by the elliptic cosine function cn. In our present paper we are interested in solutions with such tiny tails that the linear tail approximation is adequately precise. In a subsequent paper \cite{Boyd95} Boyd has presented high precision results for the case when the asymptotic phase of the tail is $\delta=0$. The use of multiple precision arithmetic allowed the calculation of extremely small tail amplitudes for $\epsilon<0.05$. We have checked that our code reproduces the tail amplitudes listed in Table II of \cite{Boyd95}. Note that since Boyd uses the variables discussed in Eq.~\eqref{equboyd}, the amplitude $\alpha$ given in that paper is equal to $6\epsilon^2\alpha$ using our notations.

The speed of our code allows us to search numerically for the phase $\delta_m$ for which the tail amplitude is minimal, $\alpha=\alpha_m$. We are not aware of such study in the literature. Using Brent's minimization method \cite{numrec07}, usually about $20$ iterations are enough to get the necessary precision. In Table \ref{tablerhoalp} we list the phase $\delta_m$ and minimal amplitude $\alpha_m$ for several choices of the parameter $\epsilon$.
\begin{table}[!hbtp]
 \centering
 \begin{tabular}{|c||c|c|c|c|}
  \hline
  $\epsilon$ & $\alpha_m$ & $\delta_m$ & $x_{\mathrm{core}}$
  & $N_{\mathrm{opt}}$\\
  \hline
  \hline
  $0.15$ & $4.1\cdot10^{-2}$ & $0.958$ & $2.56$ & $3$ \\
  \hline
  $0.1$ & $6.572\cdot10^{-4}$ & $0.60552$ & $4.77$ & $6$ \\
  \hline
  $0.07$ & $1.802403\cdot10^{-6}$ & $0.4207362$ & $7.72$ & $10$ \\
  \hline
  $0.05$ & $4.811363375\cdot10^{-10}$ & $0.3001268310$ & $12.1$ & $14$  \\
  \hline
  $0.035$ & $1.472008979\cdot10^{-15}$ & $0.2100205651$ & $18.2$ & $21$  \\
  \hline
  $0.025$ & $4.771438977\cdot10^{-23}$ & $0.1500037632$ & $26.8$ & $30$  \\
  \hline
  $0.017$ & $1.527829748\cdot10^{-35}$ & $0.1020005424$ & $41.2$ & $45$  \\
  \hline
  $0.012$ & $5.935328843\cdot10^{-52}$ & $0.0720000947$ & $60.1$ & $64$  \\
  \hline
 \end{tabular}
 \caption{Numerically calculated values of the minimal tail amplitude $\alpha_m$, the corresponding phase $\delta_m$, and the core radius $x_{\mathrm{core}}$. The order of optimal truncation for the expansion describing the core is listed in the last column (see Sec.~\ref{secexpcore}).}
 \label{tablerhoalp}
\end{table}
In the table we also include the radius of the core, which we estimate by the lowest value of $x$ where there is a zero crossing in the function $u$. For $x>x_{\mathrm{core}}$ the oscillating tail dominates. For higher $\epsilon$ values we can reach fewer digits of precision because the matching to the linear approximation of the tail brings an error of order $\alpha_m^2$. In these cases there is a slight dependence on how the outer boundary $L$ is chosen. For $\epsilon<0.05$ we present only $10$ digits, even if we can reach higher precision  with our spectral code. The reason for such precise calculations is that we intend to compare to the higher order analytical results presented in the next sections. The smaller $\epsilon$ is the more computational resources are necessary. To get the values for $\epsilon=0.012$ we have used $5000$ collocation points with $105$ digits arithmetic, and the calculation using the ARB library took several hours on a desktop computer.

\section{Expansion procedure for the core} \label{secexpcore}

From this section on we concentrate on analytical methods and compare them to our numerical results.
We construct an asymptotic expansion, in powers of the small parameter $\epsilon$, which can be used to describe the core region of an almost localized solitary wave solution. Since the amplitude of the oscillating tail in the faraway region is exponentially small in $\epsilon$, this expansion is not able to describe those oscillations. Into Eq.~\eqref{equxxxx} we substitute the expansions
\begin{align}
 u&=\sum_{n=0}^{\infty}u_n\epsilon^{2n} \,, \label{equuneps} \\
 c&=\sum_{n=0}^{\infty}c_n\epsilon^{2n} \,,
\end{align}
where $u_n$ are functions of $x$ and $c_n$ are numbers \cite{Grimshaw95,Boyd98}. To obtain a meaningful finite result, the asymptotic series has to be truncated at some positive integer order. The error of the approximation is the smallest when the series is truncated at the optimal order $N_{\mathrm{opt}}$, which is expected to increase proportionally with $1/\epsilon$. The error of the optimally truncated series is anticipated to be exponentially small in $\epsilon$, just as the oscillating tail that this expansion cannot describe. Substituting into Eq.~\eqref{equxxxx}, the vanishing of the $\epsilon$ independent part gives the KdV equation
\begin{equation}
 u_{0,xx}+3u_0^2-c_0u_0=0 \,,
\end{equation}
which has the solution given by \eqref{equ0twogamsq}. For a given $c_0$ this is the unique localized single-core solution when symmetry with respect to $x=0$ is assumed. For $n>0$, the vanishing of the coefficient of $\epsilon^{2n}$ yields
\begin{equation}
 u_{n-1,xxxx}+u_{n,xx}+\sum_{j=0}^{n}(3u_j-c_j)u_{n-j}=0 \,.
\end{equation}
If we assume that the functions are known up to order $n-1$, then this equation can be considered as a linear inhomogeneous differential equation for determining $u_n$,
\begin{equation}
 u_{n,xx}+6u_0u_n-c_0u_n = R_n \,, \label{equnxxrrn}
\end{equation}
where
\begin{equation}
 R_n=-u_{n-1,xxxx}-3\sum_{j=1}^{n-1}u_{j}u_{n-j}
 +\sum_{j=1}^{n}c_{j}u_{n-j} \,. \label{eqrrnuxxxx}
\end{equation}
General solutions of the homogeneous problem with $R_n=0$ can be obtained as linear combinations of two solutions. The first solution is the derivative of $u_0$, which is antisymmetric. The other solution blows up exponentially at infinity. It follows that the solution of the inhomogeneous problem \eqref{equnxxrrn} which is symmetric with respect to $x=0$ and localized has to be unique.

Proceeding order by order in $n$, it turns out that the inhomogeneous source term can be written as a finite sum of powers of $\mathrm{sech}^{2}(\gamma x)$,
\begin{equation}
 R_n=\sum_{j=1}^{n+2}R_{n,j}\gamma^{2n+4}\,\mathrm{sech}^{2j}(\gamma x) \,.
\end{equation}
The functions $u_n$ can be expanded similarly,
\begin{equation}
 u_n=\sum_{j=1}^{n+1}u_{n,j}\gamma^{2n+2}\,\mathrm{sech}^{2j}(\gamma x)
 \,. \label{eqununksech}
\end{equation}
The powers of $\gamma$ are included in order to make $R_{n,j}$ and $u_{n,j}$ rational numbers without $\gamma$ factors. Comparing with \eqref{equ0twogamsq} follows that $u_{0,1}=2$.

Using the identities for the derivatives of $\mathrm{sech}^{j}x$ (see, e.g., Appendix A of \cite{Boyd98}), from \eqref{eqrrnuxxxx} it follows that for $n\geq 1$,
\begin{align}
 R_{n,j}=&-16j^4u_{n-1,j}+8(j-1)(2j-1)(2j^2-2j+1)u_{n-1,j-1} \notag \\
 &-(2j-4)(2j-3)(2j-2)(2j-1)u_{n-1,j-2} \label{eqrrnjm16}\\
 &-3\sum_{l=1}^{n-1}\sum_{m=1}^{l+1}u_{l,m}u_{n-l,j-m}
 +\sum_{l=1}^{n-j+1}\hat c_lu_{n-l,j} \,, \notag
\end{align}
where $\hat c_l=\gamma^{-2l-2}c_l$. This expression for $R_{n,j}$ is valid only if we substitute zero for every occurrence of $u_{n,j}$ when $j<1$ or $j>n+1$. If the coefficients $u_{n,j}$ are known up to order $n-1$ in the first index, then \eqref{eqrrnjm16} can be used to calculate the source term $R_n$ in \eqref{equnxxrrn}. For $j\geq2$ all $\mathrm{sech}^{2j}(\gamma x)$ terms in $R_n$ can be generated from appropriate sech terms in $u_n$. However, when setting the right-hand side of \eqref{equnxxrrn} to $\mathrm{sech}^{2}(\gamma x)$ the symmetric, asymptotically decaying solution can be written as
\begin{equation}
 u_n=\frac{1}{2}\,\mathrm{sech}^2(\gamma x)(1-\gamma x\tanh(\gamma x))  \,.
\end{equation}
This solution goes to zero at infinity as $x\exp(-2\gamma x)$ which is a slower decay than the $\exp(-2\gamma x)$ decay of the other terms. Because of this, we must avoid this source term by setting $R_{n,1}=0$ for all $n$. Since $R_{n,1}$ contains $\hat c_n$ linearly, this can be achieved at any order by the appropriate choice of $\hat c_n$. Proceeding with the calculation it turns out that $\hat c_0=4$, $\hat c_1=16$ and $\hat c_j=0$ for all $j\geq 2$, consistently with \eqref{eqc4gamp}.

From \eqref{equnxxrrn} it follows that for $2\leq j\leq n+1$
\begin{equation}
 4(j^2-1)u_{n,j}+\left[12-(2j-1)(2j-2)\right]u_{n,j-1}=R_{n,j} \,. \label{equnjunjp1}
\end{equation}
For $j=n+2$ we obtain
\begin{equation}
 \left[12-(2n+3)(2n+2)\right]u_{n,n+1}=R_{n,n+2} \,. \label{equnnp1rr}
\end{equation}
The equation for $j=1$ is simply $R_{n,1}=0$. If all $R_{n,j}$ coefficients are already calculated at order $n$, then $u_{n,n+1}$ can be obtained from \eqref{equnnp1rr}. After this, all $u_{n,j-1}$ can be calculated one by one in decreasing order in $j$ using \eqref{equnjunjp1}. This algorithm can easily be implemented using any algebraic manipulation software. Results for the coefficients up to order $n=4$ are given in Table \ref{tableunj}.
\begin{table}[!hbtp]
 \centering
 \begin{tabular}{|c||c|c|c|c|c|}
  \hline
  $n\downarrow\ j\rightarrow$ & $1$ & $2$ & $3$ & $4$ & $5$ \\
  \hline
  $0$ & $2$ & - & - & - & - \\
  $1$ & $-20$ & $30$ & - & - & - \\
  $2$ & $60$ & $-930$ & $930$ & - & - \\
  $3$ & $-2472$ & $21036$ & $-66216$ & $49662$ & - \\
  $4$ & $-\dfrac{240780}{7}$ & $-\dfrac{3177030}{7}$ & $\dfrac{23319570}{7}$
  & $-\dfrac{48197250}{7}$ & $\dfrac{28918350}{7}$ \\
  \hline
 \end{tabular}
 \caption{First few values of $u_{n,j}$.}
 \label{tableunj}
\end{table}
A similar algorithm to calculate $u_{n,j}$ has been presented in Table 10.5 of Boyd's book \cite{Boyd98}. Note that there is a typo in the pseudocode there: the summation in the phase speed contributions should start from $m=1$, not from $0$.

\begin{figure}[!hbt]
 \centering
 \includegraphics[width=115mm]{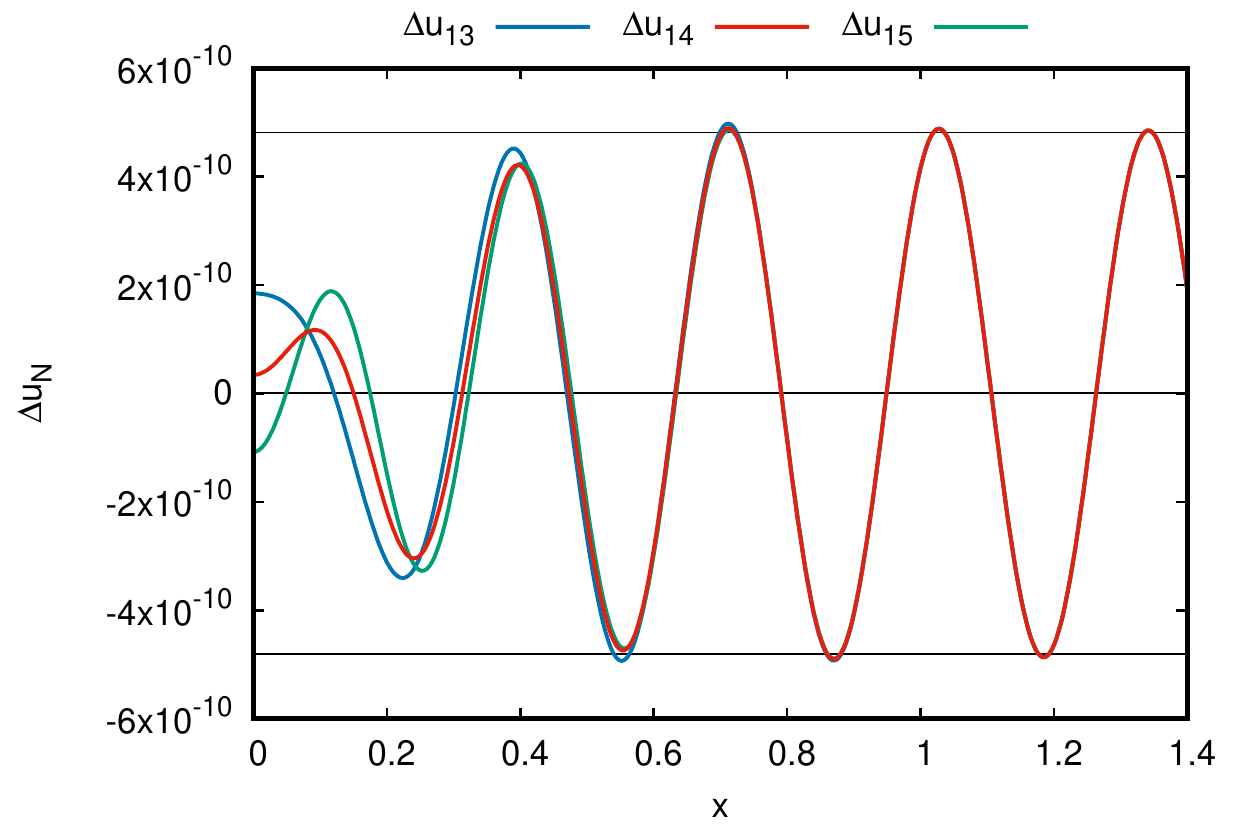}
\caption{\label{figepserr} Difference of the $N$th order approximation of the $\epsilon$ expansion \eqref{equuneps} from the minimal tail symmetric solution $u$ for $\epsilon=0.05$. The horizontal lines show the tail amplitude $\alpha_m=4.811363375\cdot10^{-10}$.}
\end{figure}

If the solution $u$ has already been calculated precisely by some numerical method, we can compare it to various orders of the above $\epsilon$ expansion, defining the error of the $N$th order analytic approximation by
\begin{equation}
 \Delta u_N^{}=u-\sum_{n=0}^{N}u_n\epsilon^{2n} \,.
\end{equation}
Choosing $\epsilon=0.05$ and the symmetric solution $u$ with the minimal tail, in Fig.~\ref{figepserr}
we plot the functions $\Delta u_N^{}$ for three values of $N$ for which the error is the smallest. Clearly, for this $\epsilon$ value the optimal truncation is at $N_{\mathrm{opt}}=14$. Since all $u_n$ decay exponentially, for large $x$ the difference should agree with the oscillating tail. The figure shows that for the optimal truncation this holds even in most of the inner region, since now the core radius is $x_{\mathrm{core}}=12.1$. It turns out that close to the center the error is even smaller. We have obtained similar plots for the other $\epsilon$ values listed in Table \ref{tablerhoalp}. The values of $N_{\mathrm{opt}}$ for various $\epsilon$ values are listed in the last column of the table. It can be checked that $N_{\mathrm{opt}}$ increases proportionally to $1/\epsilon$. The contribution of the $n$th term in the expansion \eqref{equuneps}, i.e., the function $u_n\epsilon^{2n}$, is generally the smallest for $n=N_{\mathrm{opt}}+1$, or in some cases for $n=N_{\mathrm{opt}}$, as one can expect it for asymptotic series.

\section{WKB solution} \label{secexpcorr}

We intend to linearize Eq.~\eqref{equxxxx} around some solution $u$. This solution may be a numerically obtained symmetric solution with a small tail in both directions, or an asymmetric solution that tends to zero for positive $x$. The important point is that the core region should be approximated well by the expansion \eqref{equuneps}. Substituting $u\to u+u_w$ into \eqref{equxxxx} and linearizing gives
\begin{equation}
 \epsilon^2u_{w\,xxxx}+u_{w\,xx}+6u u_w-cu_w = 0 \,. \label{equwxxxx}
\end{equation}
We will apply this formalism for $u_w$ which turns out to be exponentially small in terms of $\epsilon$; hence the linear approximation is well justified.
As we have seen, in the asymptotic region there are oscillations with spatial frequency proportional to $1/\epsilon$. Hence we use the WKB method to search for solutions of \eqref{equwxxxx}. The first step is to substitute $u_w=\beta_c\exp A$, where $A$ is a function of $x$, and $\beta_c$ is a complex constant. Then we expand $A$ in powers of $\epsilon$, starting with a $1/\epsilon$ term,
\begin{equation}
 A=\sum_{n=-1}^\infty A_n\epsilon^n\,,
\end{equation}
and solve the resulting equation order by order in $\epsilon$. Since all $A_k$ appear only in differentiated form, there will be additive complex scalar freedom in all of these functions. All these can be absorbed into the complex valued $\epsilon$ dependent factor $\beta_c$.

To leading $\epsilon^{-2}$ order we obtain
$\left(A_{-1\, x}\right)^2\left[\left(A_{-1\, x}\right)^2+1\right]=0$.
Since we are looking for high frequency solutions we are not interested in the $A_{-1\, x}=0$ solution. We continue with the choice $A_{-1\, x}=-i$, since the solution obtained from $A_{-1\, x}=i$ turns out to be the complex conjugate to all orders in $\epsilon$. Proceeding order by order in $\epsilon$, at each order we obtain a condition determining $A_{n\, x}$. The first few functions are
\begin{align}
 A_{-1}&=-ix \,, \\
 A_0&=0 \,, \\
 A_1&=-2i\gamma^2 x+6i\gamma\tanh(\gamma x) \,, \\
 A_2&=15\gamma^2\mathrm{sech}^2(\gamma x) \,, \\
 A_3&=2i\gamma^4 x+111i\gamma^3\mathrm{sech}^2(\gamma x)\tanh(\gamma x)
  \,, \\
 A_4&=\frac{525}{2}\gamma^4\mathrm{sech}^2(\gamma x)
 \left[3\mathrm{sech}^2(\gamma x)-2\right] \,, \\
 A_5&=-4i\gamma^6 x+\frac{3}{5}i\gamma^5\left[
 12267\mathrm{sech}^4(\gamma x)-4089\mathrm{sech}^2(\gamma x)+632
 \right]\tanh(\gamma x)
  \,, \\
 A_6&=\frac{3}{2}\gamma^6\mathrm{sech}^2(\gamma x)
 \left[49317\mathrm{sech}^4(\gamma x)-49317\mathrm{sech}^2(\gamma x)+8050\right] \,.
\end{align}
It is natural to choose the value of the additive constants in $A_{2n-1}$ to make the functions antisymmetric at $x=0$. The form of $A_{2n}$ can be made unique by requiring that the functions tend to zero at infinity.

The terms proportional to $x$ in the odd indexed $A_n$ functions can be absorbed into the $A_{-1}\epsilon^{-1}=-ix/\epsilon$ term if we replace it by $-ikx/\epsilon$, where $k$ is the $\epsilon$ dependent constant defined in \eqref{eqksqrt}. This can be done by setting
\begin{equation}
 -\frac{ix}{\epsilon}+\sum_{\substack{n=1\\ \mathrm{odd}}}^{\infty}
 A_{n}\epsilon^{n}
 =-\frac{ikx}{\epsilon}+\sum_{\substack{n=1\\ \mathrm{odd}}}^{\infty}
 i\tilde A_{n}\epsilon^{n} \,, \label{eqixanikxan}
\end{equation}
where for positive integer $n$ the functions $\tilde A_{2n-1}$ are defined by
\begin{equation}
 A_{2n-1}=i\tilde A_{2n-1}-ix\gamma^{2n}\frac{(-1)^{n+1}(2n)!}
 {(2n-1)(n!)^2} \,.
\end{equation}
Inserting the $k$ factor into the linear term is natural, since the asymptotic spatial frequency is $k/\epsilon$, which is valid to all orders in $\epsilon$.
All these new functions with odd indices have a finite limit at infinity, which we denote by \begin{equation}
\tilde\delta_{2n-1}=\lim_{x\to+\infty}\tilde A_{2n-1} \,, \label{eqtildelta}
\end{equation}
since they will determine the asymptotic phase shift of the minimal tail configuration. The first few values are
\begin{align}
 \tilde\delta_1&=6\gamma \,, \label{eqtildel1}\\
 \tilde\delta_3&=0 \,, \\
 \tilde\delta_5&=\frac{1896}{5}\gamma^5 \,, \\
 \tilde\delta_7&=\frac{67140}{7}\gamma^7 \,, \\
 \tilde\delta_9&=\frac{2662320}{7}\gamma^9 \,, \\
 \tilde\delta_{11}&=\frac{1301363652}{77}\gamma^{11} \,.
\end{align}
We have calculated $\tilde\delta_n$ for $n<100$ using algebraic manipulation software.

For real $x$ all even indexed $A_n$ are real, and hence they will contribute to the amplitude of the linearized solution. The odd $A_n$ are all purely imaginary, so they will determine the phase. The general solution of the linearized problem that takes real values for real $x$ can be obtained by a linear combination of the solutions belonging to $A_{-1}=-ix$ and $A_{-1}=ix$,
\begin{equation}
 u_w=\beta\exp\left(\sum_{\substack{n=2\\ \mathrm{even}}}^{\infty}
 A_n\epsilon^n\right)
 \sin\left(\frac{kx}{\epsilon}-\delta_w
 -\sum_{\substack{n=1\\ \mathrm{odd}}}^{\infty}
 \tilde A_n\epsilon^n\right) \,. \label{equwexp}
\end{equation}
Here $\beta$ and $\delta_w$ are real constants with arbitrary $\epsilon$ dependence. They are related to the magnitude and phase of the complex constant $\beta_c$. The exponential term can be directly expanded in powers of $\epsilon$, providing
\begin{equation}
 u_w=\beta\left(1+\sum_{\substack{n=2\\ \mathrm{even}}}^{\infty}
 \tilde A_n\epsilon^n\right)
 \sin\left(\frac{kx}{\epsilon}-\delta_w
 -\sum_{\substack{n=1\\ \mathrm{odd}}}^{\infty}
 \tilde A_n\epsilon^n\right) \,, \label{equwnoexp}
\end{equation}
where the even indexed coefficients $\tilde A_n$ can easily be obtained from the original $A_n$. The first few functions are $\tilde A_2=A_2$, $\tilde A_4=A_4+\frac{1}{2}A_2^{\,2}$ and $\tilde A_6=A_6+A_4A_2+\frac{1}{6}A_2^{\,3}$. For $\delta_w=0$ the function $u_w$ is antisymmetric with respect to $x=0$, while for $\delta_w=\pi/2$ it is symmetric.

\subsection{Phase of the tail}

Since the $A_{2n}$ functions tend to zero at infinity, $\beta$ gives the asymptotic amplitude of the oscillation represented by $u_w$ in \eqref{equwnoexp}. The amplitude in the core region is modified by the factor which is $O(1)$, so the linear correction $u_w$ remains small even in the core region. The asymptotic behavior of the function for positive $x$ is $u_w=\beta\sin(kx/\epsilon-\delta_w-\delta_m)$ with
\begin{equation}
 \delta_m=\sum_{\substack{n=1\\ \mathrm{odd}}}^{\infty}
 \tilde\delta_n\epsilon^n \,, \label{eqdeltildeexp}
\end{equation}
where the constants $\tilde\delta_n$ for odd $n$ are defined in \eqref{eqtildelta}. The same function near the center behaves as $u_w\sim\sin(kx/\epsilon-\delta_w)$, since $\tilde A_{2n-1}=0$ at $x=0$ in \eqref{equwnoexp}. The asymptotic phase comes from two contributions. The part $\delta_w$ gives the phase near the center, while the $\epsilon$ dependent constant $\delta_m$ gives the additional phase shift between the center and positive infinity.

Another way to consider $u_w$ in \eqref{equwnoexp} is to decompose it to a sine part with $\delta_w=0$ and a cosine part corresponding to $\delta_w=\pi/2$, both with arbitrary amplitudes,
\begin{equation}
 u_w=\left(1+\sum_{\substack{n=2\\ \mathrm{even}}}^{\infty}
 \tilde A_n\epsilon^n\right)\left[
 \beta_{\sin}
 \sin\left(\frac{kx}{\epsilon}
 -\sum_{\substack{n=1\\ \mathrm{odd}}}^{\infty}
 \tilde A_n\epsilon^n\right)
 +\beta_{\cos}
 \cos\left(\frac{kx}{\epsilon}
 -\sum_{\substack{n=1\\ \mathrm{odd}}}^{\infty}
 \tilde A_n\epsilon^n\right)
 \right] .
\end{equation}
The sine part is antisymmetric at the center $x=0$, while the cosine part is symmetric. The cosine part with arbitrary $\beta_{\cos}$ amplitude can be added to any symmetric solution $u$ of \eqref{equxxxx}, showing that the symmetric solution  is not unique. On the other hand, the sine part of the tail of any symmetric solution has a fixed amplitude. Let us suppose that a certain symmetric solution $u$ of \eqref{equxxxx} has a small-amplitude tail given by \eqref{equtail} at large distances for $x>0$. This tail can also be written as
\begin{equation}
 u=\alpha\cos\left(\delta-\delta_m\right)
 \sin\left(\frac{kx}{\epsilon}-\delta_m\right)
 -\alpha\sin\left(\delta-\delta_m\right)
 \cos\left(\frac{kx}{\epsilon}-\delta_m\right)\,. \label{equscossin}
\end{equation}
Adding $u_w$ with $\beta_{\sin}=0$ and $\beta_{\cos}=\alpha\sin\left(\delta-\delta_m\right)$, the cosine part of the tail becomes completely canceled, and we obtain the minimal tail symmetric solution, $u_m=u+u_w$. Adding $u_w$ with nonzero $\beta_{\sin}$ would destroy the central symmetry of $u$. For given $\epsilon$ the minimal tail solution is unique, and the tail for $x>0$ has the asymptotic behavior
\begin{equation}
 u_m=\alpha_m\sin\left(\frac{kx}{\epsilon}-\delta_m\right)\,, \label{equmasympt}
\end{equation}
where $\alpha_m=\alpha\cos\left(\delta-\delta_m\right)$.
Consequently, the minimal tail-amplitude symmetric solution necessarily has the  phase $\delta_m$ in the tail, which has already been calculated in terms of the asymptotic series \eqref{eqdeltildeexp}.

\begin{figure}[!hbt]
 \centering
 \includegraphics[width=115mm]{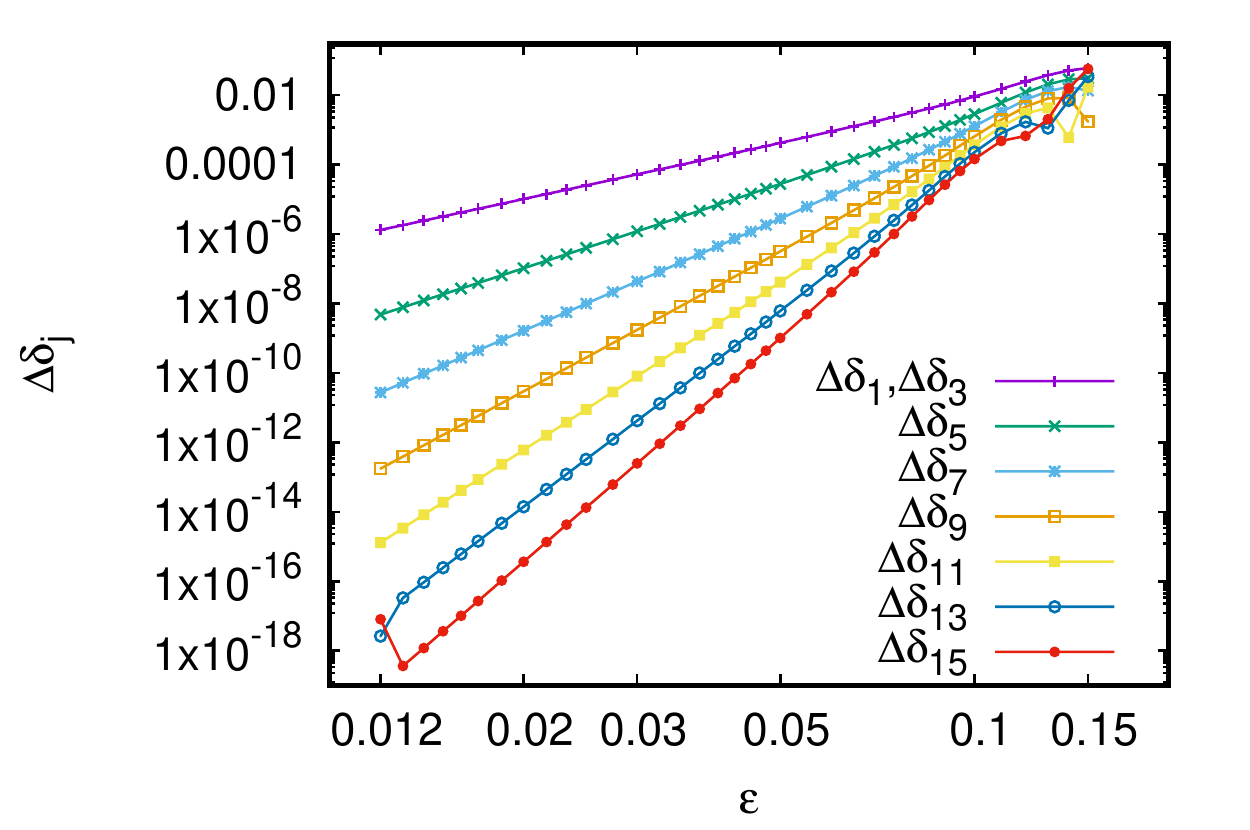}
\caption{\label{figdediff} Log-log plot of $\Delta\delta_j=|(\delta_m-\delta_m^{(j)})/\delta_m|$, showing the relative difference of the numerically calculated $\delta_m$ from its various order analytic approximations $\delta_m^{(j)}$. }
\end{figure}

In Fig.~\ref{figdediff} we compare the numerically calculated minimal-amplitude asymptotic phase $\delta_m$ and its various order approximations provided by \eqref{eqdeltildeexp},
\begin{equation}
 \delta_m^{(j)}=\sum_{\substack{n=1\\ \mathrm{odd}}}^{j}
 \tilde\delta_n\epsilon^n \,.
\end{equation}
We plot logarithmically the relative difference $\Delta\delta_j=|(\delta_m-\delta_m^{(j)})/\delta_m|$ for $j\leq 15$. Since $\tilde\delta_3=0$, naturally $\Delta\delta_1=\Delta\delta_3$. For $\epsilon\lesssim 0.07$ the numerical values of $\Delta\delta_j$ are decreasing as $\epsilon^{j+1}$. The deviations from the straight line of $\tilde\delta_{13}$ and $\tilde\delta_{15}$ for $\epsilon=0.012$ show that the numerical result for this $\epsilon$ value is less than $18$ digits precise. This is due to the necessary high resolution and long running times for such small values of $\epsilon$. Although we do not show $\Delta\delta_j$ for higher $j$ in the figure, the precision of this agreement can be further improved for not so small $\epsilon$ values. For example, for $\epsilon=0.02$ our numerically calculated $\delta_m$ agrees to $34$ digits with $\delta_m^{(47)}$. The fact that we obtain numerical results correct to so many digits of precision shows the remarkable power of the exponentially convergent spectral method when it is combined with arbitrary precision arithmetic. On the other hand, for high $\epsilon$ values it is apparent that \eqref{eqdeltildeexp} is indeed an asymptotic series. For example, for $\epsilon=0.15$ the best approximation of the numerical result is given by $\delta_m^{(9)}$, higher orders giving larger and larger errors. In this case the smallest $\tilde\delta_n\epsilon^n$ contribution to the sum in \eqref{eqdeltildeexp} belongs to $n=11$, which is just a little smaller than the contribution of the $n=9$ term. In general, an asymptotic series is expected to give the best approximation when the summing is stopped at the term that gives the smallest contribution to the result.

Another important consequence of \eqref{equscossin} is that for a given $\epsilon$ the tail amplitude $\alpha$ of any symmetric solution with phase $\delta$ is related to the minimal tail amplitude $\alpha_m$ by
\begin{equation}
 \alpha=\frac{\alpha_m}{\cos\left(\delta
 -\delta_m\right)} \,. \label{eqalpsalpm}
\end{equation}
Since $\alpha_m$ is exponentially small, this relation is valid to higher polynomial orders in $\epsilon$. To numerically check Eq.~\eqref{eqalpsalpm}, for various $\epsilon$ values we have calculated $\delta_m$, $\alpha_m$, and for different $\delta$ phase shifts the amplitudes $\alpha$. According to our results, the error in \eqref{eqalpsalpm} turns out to be order $\alpha^2$, similar to the error in the numerically calculated $\alpha$ caused by the linear tail approximation. Because of the high precision of this relation it is appropriate to concentrate on the minimal tail amplitude in the following.
We should like to point out that in the limit of $\delta\to\pi/2$, Eq.\ \eqref{eqalpsalpm} yields
\begin{equation} \label{eqalpsalpm1}
\alpha\vert^{}_{\delta=\frac{\pi}{2}}=\alpha_m\left(\frac{1}{6\gamma\epsilon}+{\cal{O}}(\epsilon)\right) \,,
\end{equation}
which agrees with the result of Ref.\ \cite{Sun98}.
Taking $u_m$ as the minimal tail symmetric solution with asymptotics in \eqref{equmasympt} and subtracting $u_w$ with $\beta=\alpha_m$ and $\delta_w=0$ in \eqref{equwnoexp}, the oscillating tail of the asymmetric solution $u_{-}=u_m-u_w$ becomes totally canceled in the positive $x$ direction. Using the linear approximation, the solution $u_{-}$ would appear to have double tail amplitude in the negative directions. However, it has been shown in \cite{Grimshaw95} using an energy flux conservation law, that no such solution can exist. From numerical simulations we can actually see that although the core domain remains quite similar to that of the symmetric one, the asymmetric solution blows up in the negative $x$ direction before the double-amplitude tail could appear. Nevertheless, $u_{-}$ plays an important role in the analytical calculation of the minimal tail amplitude.

\section{Asymptotic matching on the complex plane}\label{sec:match}

\subsection{Complex extension}

Since the amplitude $\alpha$ of the tail is exponentially small in terms of $\epsilon$, it cannot be determined by the direct expansion and WKB methods that we have used for the calculation of the phase $\delta$. To deal with the amplitude we have to extend the functions to the complex $x$ plane and study them near the singularity closest to the real $x$ axis, applying the method of matched asymptotic expansions \cite{SegurKruskal87,Pomeau88}. We will apply the Laplace transform to solve the inner problem \cite{Grimshaw95}.

We extend analytically Eq.~\eqref{equxxxx} and its solution $u$ to the complex $x$ plane. The value of the function $u$ at the point $x=x_r-ix_i$ has to be the complex conjugate of the value at $x=x_r+ix_i$, according to the Schwarz reflection principle. If the original real function is symmetric at $x=0$, then the extension will naturally satisfy $u(x_r+ix_i)=u(-x_r-ix_i)$. In this case the value of $u$ at $x=-x_r+ix_i$ will be the complex conjugate of the value at $x=x_r+ix_i$, and the function must take real values on the imaginary axis. The extension of an antisymmetric function, satisfying $u(-x)=-u(x)$, has the property that $u(-x_r+ix_i)$ is $-1$ times the complex conjugate of $u(x_r+ix_i)$, and it must be purely imaginary on the imaginary axis.

We also extend to the complex $x$ plane the asymptotic expansion \eqref{equuneps} of $u$, where $u_n$ depends on $x$ according to \eqref{eqununksech}, with already calculated coefficients $u_{nj}$. All $x$ dependence is through powers of the function $\mathrm{sech}^2(\gamma x)$, which is singular at the points $x=(2n+1)i\pi/(2\gamma)$, for integer $n$. The closest singularity above the real axis is at $x=i\pi/(2\gamma)$. In a neighborhood of this singularity we define the rescaled $q$ complex coordinate by
\begin{equation}
 x=\frac{i\pi}{2\gamma}+\epsilon q \,. \label{eqxwithq}
\end{equation}
Near the singularity $\mathrm{sech}^2(\gamma x)$ has a Laurent series expansion starting with a $\epsilon^{-2}q^{-2}$ term. The expansion of $u$ can be obtained by adding various powers of this using some algebraic manipulation software. Since $u$ is growing as $\epsilon^{-2}$ near the singularity, it is natural to write the result in terms of the rescaled function
\begin{equation}
 v=\epsilon^2 u \,. \label{eqvepsepsu}
\end{equation}
Substituting \eqref{eqxwithq} into \eqref{equuneps}, expanding in powers of $1/q$, and then also in powers of $\epsilon$, we obtain that
\begin{equation}
 v=\sum_{n=0}^{\infty}\gamma^{2n}\epsilon^{2n}v_n \,, \label{eqvvnepsnn}
\end{equation}
where the expansions of the first three functions are
\begin{align}
 v_0&=-\frac{2}{q^2}+\frac{30}{q^4}
 -\frac{930}{q^6}+\frac{49662}{q^8}
 -\frac{28918350}{7q^{10}}+\ldots \label{eqv0qexp}\\
 v_1&=\frac{2}{3} \label{eqv1qeps}\\
 v_2&=-\frac{2q^2}{15}+\frac{2}{3}
 +\frac{64}{q^2}+\frac{5856}{5q^4}
 -\frac{827520}{7q^6}+\ldots \,. \label{eqv2qexp}
\end{align}
Since there are only even powers of $q$ with real coefficients, any truncated versions of the above series correspond to the complex extension of symmetric functions. Apart from the exactly known constant $v_1$ all $v_n$ functions are given in terms of asymptotic expansions in $1/q$. We use these expansions in some domain where both $\epsilon q$ and $1/q$ are small; hence $\epsilon^{2n}v_n$ can be small too.

\subsection{Inner problem}

Using the rescaled function $v$ and the complex coordinate $q$ Eq.~\eqref{equxxxx} becomes
\begin{equation}
 v_{qqqq}+v_{qq}+3v^2-\epsilon^2 cv = 0 \,, \label{eqvinnerqqqq}
\end{equation}
where $c$ also depends on $\epsilon$ according to \eqref{eqc4gamp}. If we search for solutions of this equation as an expansion of the form \eqref{eqvvnepsnn}, then taking the various $\epsilon^n$ contributions we obtain differential equations for $v_n$. The first two equations are:
\begin{align}
 &v_{0\,qqqq}+v_{0\,qq}+3v^2_0 = 0 \,, \label{eqv0qqqq}\\
 &v_{1\,qqqq}+v_{1\,qq}+2v_0(3v_1-2) = 0 \,. \label{eqv1qqqq}
\end{align}
The equation for $v_2$ will be studied in Sec.~\ref{secfourth}. Independently of $v_0$, the second equation can always be solved by $v_1=2/3$. Since this also agrees with the result obtained in \eqref{eqv1qeps}, we will use this solution for $v_1$ from now on.
These equations for $v_n$ are the $n$th order representations of the so-called inner problem. The expansion solutions \eqref{eqv0qexp}-\eqref{eqv2qexp} of the outer problem will be used as matching conditions for $v_n$. These asymptotic expansions provide valid boundary conditions when $\mathrm{Im}\,q<0$ is fixed and $\mathrm{Re}\,q\to+\infty$, determining unique inner solutions, which we denote by $v_n^{(-)}$. The functions $v_n^{(-)}$ do not have any oscillating tail in the positive direction $\mathrm{Re}\,q>0$. They can be associated with the unique asymmetric solution $u_{-}$ of the original Eq.~\eqref{equxxxx} for which there is no tail in the positive $x$ direction.

\subsubsection{Zeroth order inner problem}

An appropriate precise solution of the $\epsilon$ independent Eq.~\eqref{eqv0qqqq} can be used to determine the tail amplitude of the solution $u$ of \eqref{equxxxx} to leading order in $\epsilon$. An expansion solution for $v_0$ consistent with \eqref{eqv0qexp} can be searched for in the form
\begin{equation}
 v_0=\sum_{n=1}^{\infty}b^{(0)}_n q^{-2n} \,. \label{eqv0expq}
\end{equation}
Equation \eqref{eqv0qqqq} gives the following equation for the coefficients:
\begin{equation}
 (2n-2)(2n-1)(2n)(2n+1)b^{(0)}_{n-1}+(2n)(2n+1)b^{(0)}_n
 +3\sum_{j=1}^{n}b^{(0)}_{j}b^{(0)}_{n-j+1}=0 \,. \label{eqbncf1}
\end{equation}
For $n=1$ there is no $b^{(0)}_{n-1}$ term, and the nonzero solution is $b^{(0)}_1=-2$. For $n\geq 2$ we can obtain a recursion relation by taking the first and the last terms out of the summation,
\begin{equation}
 (2n-3)(2n+4)b^{(0)}_n=-(2n-2)(2n-1)(2n)(2n+1)b^{(0)}_{n-1} -3\sum_{j=2}^{n-1}b^{(0)}_{j}b^{(0)}_{n-j+1} \,. \label{eqbn0rec}
\end{equation}
The obtained coefficients are the same as those in \eqref{eqv0qexp}. For large $n$ the coefficients diverge as $b^{(0)}_n\sim (-1)^n(2n-1)!\,$.

Following the method introduced in \cite{Grimshaw95}, we look for the solution of \eqref{eqv0qqqq} in the form of a Laplace transform of a function $V'_0(s)$,
\begin{equation}
 v_0=\int_\Gamma I_0(s)\mathrm{d}s \,,\quad \quad
 I_0(s)=\exp(-sq)V'_0(s) \,, \label{eqv0int}
\end{equation}
where the contour $\Gamma$ is from $s=0$ to infinity, satisfying $\mathrm{Re}(sq)>0$. The prime is here because $V'_0(s)$ is the derivative of a function $V_0(s)$ that can be defined in terms of the Borel transformation\cite{Pomeau88}. However, we will only use the $V'_0(s)$ defined by \eqref{eqv0int} in the following. Using the identity for the Laplace transform of powers of $s$ we can see that $V'_0(s)$ can be expanded as
\begin{equation}
 V'_0(s)=\sum_{n=0}^{\infty}a^{(0)}_n s^{2n+1} \,, \label{eqvvpr0}
\end{equation}
where
\begin{equation}
 a^{(0)}_n=\frac{b^{(0)}_{n+1}}{(2n+1)!} \,. \label{eqan0bnp10}
\end{equation}
This can also be written as $b^{(0)}_{n}=(2n-1)!\,a^{(0)}_{n-1}$. Substituting into \eqref{eqbncf1} we get
\begin{equation}
 a^{(0)}_{n-1}+a^{(0)}_n
 +\frac{3}{(2n+3)!}\sum_{j=0}^{n}
 (2j+1)!(2n-2j+1)!\,a^{(0)}_{j}a^{(0)}_{n-j}=0 \,. \label{eqan0cf}
\end{equation}
Separating the first and last terms in the summation and using that $a^{(0)}_{0}=-2$, one can obtain a recursion relation which is valid for $n\geq 1$,
\begin{equation}
 \frac{(n+3)(2n-1)}{(n+1)(2n+3)}a^{(0)}_n=-a^{(0)}_{n-1}
 -\frac{3}{(2n+3)!}\sum_{j=1}^{n-1}
 (2j+1)!(2n-2j+1)!\,a^{(0)}_{j}a^{(0)}_{n-j} \,. \label{eqan0rec}
\end{equation}

The singularities of the function $V'_0(s)$ will be determined by the large $n$ behavior of the coefficients. The leading order behavior is $a^{(0)}_n\approx K(-1)^n$, where $K\approx 19.97$. The constant $K$ will be very important in the following, since it will determine the radiation amplitude. Hence we intend to calculate $K$ to several digits precision. Unfortunately, no fully analytical method is known for this calculation. We look for the large $n$ behavior of the coefficients in the form
\begin{equation}
 a^{(0)}_n=(-1)^n\,\tilde a^{(0)}_n \,,\quad \qquad
 \tilde a^{(0)}_n=K G_0(n) \,,\quad \qquad
 G_0(n)=1+\sum_{j=1}^\infty\frac{g_j}{n^j} \,. \label{eqa0ngn}
\end{equation}
The constants $g_j$ should be determined by substituting this expansion into \eqref{eqan0cf}. We assume that $n$ is large, but if we are interested in a finite number of $g_j$ constants we do not have to take into account all the $n+1$ terms in the summation in \eqref{eqan0cf}. We substitute \eqref{eqa0ngn} into the equation
\begin{equation}
 \tilde a^{(0)}_n-\tilde a^{(0)}_{n-1}
 +\sum_{j=0}^{j_m}(-1)^j\, W_{n,j}^{(0)}\,
 \tilde a^{(0)}_{n-j}=0 \,, \label{eqanm1ankm}
\end{equation}
where
\begin{equation}
 W_{n,j}^{(0)}=\frac{6}{(2n+3)!}(2j+1)!
 (2n-2j+1)!\,a^{(0)}_{j} \,, \label{eqwnk0}
\end{equation}
and $j_m$ is some positive integer.

The more terms we intend to determine for $G_0(n)$ in \eqref{eqa0ngn}, the higher $j_m$ we should choose. However, the constants $W_{n,j}^{(0)}$ only involve $a^{(0)}_{j}$ with small $j$, so they can be calculated explicitly. We can use algebraic manipulation software to substitute a truncated version of the expansion \eqref{eqa0ngn} into \eqref{eqanm1ankm}. Taking the coefficients in increasing powers of $1/n$, we can determine the constants $g_j$. The first seven terms of the result yield
\begin{equation}
 G_0(n)=1-\frac{3}{n}+\frac{39}{4n^2}
 -\frac{69}{2n^3}+\frac{1929}{16n^4}-\frac{3381}{8n^5}
 +\frac{46041}{32n^6}-\frac{1089483}{224n^7}+\ldots
 \,. \label{eqa0nexpn}
\end{equation}
To get the correct coefficients up to this order one has to set at least $j_m=3$. This means that we use at least the first four and last four terms from the summation in \eqref{eqan0cf}, but can neglect the others in between. The importance of \eqref{eqa0nexpn} is that it allows us to determine the constant $K$ to several digits precision by calculating the concrete coefficients from the recursion up to moderately high $n$ values. Using some algebraic manipulation program, the calculation of $a^{(0)}_n$ by \eqref{eqan0rec} can be made faster by using floating point arithmetic valid to hundred digits precision instead of using exactly represented but very long rational numbers. In this way the first few thousand coefficients can be calculated in a couple of minutes. The approximation for the proportionality constant can be calculated using an appropriately truncated version of $G_0(n)$ as $K\approx a^{(0)}_n(-1)^n/G_0(n)$. The result up to $22$ digits is
\begin{equation}
 K=-19.96894735876096051827 \,. \label{eqkkmanydig}
\end{equation}
The high precision will be useful because we intend to study higher order $\epsilon$ corrections in the following sections.

The series \eqref{eqvvpr0} for $V'_0(s)$ is convergent for $|s|<1$ and can be analytically extended for larger $|s|$, showing that $V'_0(s)$ is unique. On the other hand, several different $v_0$ functions can be obtained from it using the Laplace transform \eqref{eqv0int}, depending on how the path $\Gamma$ is located with respect to the singularities of $V'_0(s)$. The function $V'_0(s)$ satisfies an integral equation presented in Eq.~(31) of \cite{Grimshaw95}. All singularities are located on the imaginary axis at $s=\pm n i$, where $n$ is any positive integer. There is no singularity at $s=0$.

Calculating $v_0$ by \eqref{eqv0int} in a domain where $\arg(q)$ is close to zero, it is natural to choose the positive real $s$ axis as the curve $\Gamma$. The solution obtained in this way is the asymmetric $v_0^{(-)}$, which is determined by the boundary condition \eqref{eqv0qexp} for $\mathrm{Re}\,q>0$. We can extend the $v_0^{(-)}$ defined by the Laplace transform integral to the domain $-\pi<\arg(q)\leq 0$, but for $\arg(q)\leq -\pi/2$ the contour $\Gamma$ cannot remain on the real axis, and it should move into the region where $0<\arg(s)<\pi/2$. However, if the function $v_0^{(-)}$ is defined smoothly on the positive part of the real $q$ axis, it cannot be extended by the integral \eqref{eqv0int} to the negative part of the axis, where $\arg(q)= -\pi$. That would require crossing the singularities on the upper half of the imaginary axis by the contour $\Gamma$.
The solution $v_0^{(-)}$ defined in this way will correspond to the asymmetric solution $u_{-}$ of \eqref{equxxxx} which tends to zero exponentially for $x\to\infty$ in the positive direction. This solution has a core that is very close to the core of the minimal tail symmetric solution, but it generally diverges in the negative $x$ direction. There is a similar $v_0^{(+)}$ function that can be calculated using \eqref{eqv0int}, which is valid for $-\pi\leq\arg(q)<0$. Then the contour $\Gamma$ has to be chosen in the quadrant $\pi/2<\arg(s)\leq\pi$. This solution is obviously the conjugated mirror image of the previous one with respect to the imaginary $q$ axis. The difference of $v_0^{(-)}$ and $v_0^{(+)}$ can be calculated using the residue theorem. The dominant contribution to the difference will be given by the singularity at $s=i$, so we need to determine the behavior of $V'_0(s)$ close to there. This can be inferred from the large $n$ behavior of the coefficients $a^{(0)}_n$ in \eqref{eqvvpr0}.

The residue at $s=i$ of the function $I_0(s)$ will be determined by substituting the leading order part of \eqref{eqa0ngn}, which corresponds to $G_0(n)=1$. The result can be summed,
\begin{equation}
 I_0(s)\approx\exp(-sq)K\sum_{n=0}^\infty (-1)^n s^{2n+1}=
 \exp(-sq)\frac{Ks}{1+s^2} \,. \label{eqsumexpmsq}
\end{equation}
Since the residue of $s/(1+s^2)$ is $1/2$, it follows that
\begin{equation}
 \underset{s=i}{\mathrm{Res}}\,I_0(s)=\exp(-iq)\frac{K}{2} \,.
\end{equation}
The residues at the singularities $s=ni$, where $n\geq1$ integer, will be proportional to $\exp(-inq)$, hence we can neglect their contributions. The residue theorem can be applied by choosing a curve going from $s=0$ to infinity in the domain $0<\arg(s)<\pi/2$ and coming back in the region $\pi/2<\arg(s)\leq\pi$. The difference of the two functions is
\begin{equation}
 v_0^{(-)}-v_0^{(+)}=\pi i K \exp(-iq) \,, \label{eqv0mmv0p}
\end{equation}
which is valid for any $q$ satisfying $\mathrm{Im}\,q<0$.

Grimshaw and Joshi also define a third function, $v_0^{(m)}$, by setting the contour $\Gamma$ exactly as the upper half of the imaginary $s$ axis, running through all the singularities there. This way the integral will take half of the pole contributions, yielding
\begin{equation}
 v_0^{(-)}-v_0^{(m)}=\frac{1}{2}\pi i K \exp(-iq) \,. \label{eqv0mmv0s}
\end{equation}
Since all the $a^{(0)}_n$ are real, from \eqref{eqv0int} and \eqref{eqvvpr0} follows that the function $v_0^{(m)}$ has no imaginary part on the imaginary $q$ axis; hence it corresponds to a symmetric $u$ solution on the real $x$ axis. Actually, since $K$ is real, it belongs to the one with minimal tail amplitude. Larger tail amplitude symmetric solutions could be obtained by adding $\alpha_r\exp(-iq)$ to $v_0^{(-)}$ with arbitrary real $\alpha_r$. Taking the imaginary part of \eqref{eqv0mmv0s} on the lower half of the imaginary axis we obtain that
\begin{equation}
 \mathrm{Im}\,v_0^{(-)}=\frac{1}{2}\pi K \exp(-iq) \quad
 \mathrm{for} \ \  \mathrm{Re}\,q=0 \,, \
 \mathrm{Im}\,q<0 \,. \label{eqimv0m}
\end{equation}
The behavior of $\mathrm{Im}\,v_0^{(-)}$ along the lower part of the imaginary axis for large $|q|$ is determined by the constant $K$ given in \eqref{eqkkmanydig}. This result will allow us to determine the minimal tail amplitude for symmetric solutions in the next sections.

\subsection{Complex extension of the linear correction} \label{secexpsmall}

All functions $A_n$ defined in Sec.~\ref{secexpcorr} contain various powers of $\mathrm{sech}(\gamma x)$; hence the linear correction $u_w$ is singular at the same places as the original $u$ solution. Close to the singularity at $x=i\pi/(2\gamma)$ we substitute \eqref{eqxwithq} for $x$ into \eqref{equwexp} to obtain a function depending on $q$. Expressing the sine function as a difference of two exponentials, we can neglect the small term proportional to $\exp[-k\pi/(2\gamma\epsilon)]$; hence
\begin{equation}
 u_w=\frac{i\beta}{2}\exp\left(\sum_{\substack{n=2\\ \mathrm{even}}}^{\infty}
 A_n\epsilon^n\right)
 \exp\left(\frac{k\pi}{2\gamma\epsilon}-ikq+i\delta_w
 +i\sum_{\substack{n=1\\ \mathrm{odd}}}^{\infty}
 \tilde A_n\epsilon^n\right) \,.
\end{equation}
Using \eqref{eqixanikxan} to bring back the original $A_n$, we can write this into the simpler form, using a single summation for all $n$,
\begin{equation}
 u_w=\frac{i\beta}{2}\exp\left(\frac{k\pi}{2\gamma\epsilon}\right)
 \exp\left(-iq+i\delta_w\right)
 \exp\left(-\frac{(k-1)\pi}{2\gamma\epsilon}
 +\sum_{n=1}^{\infty}
 A_n\epsilon^n\right) \,.
\end{equation}
We keep $k$ as it is in the first exponential term, but substituting the form \eqref{eqksqrt} of $k$ into the third term we see that $(k-1)/\epsilon$ is small. Hence we proceed first by expanding the argument of the third exponential in powers of $1/q$, and then we expand the exponential of the result in $\epsilon$. The result correct up to order $\epsilon^4$ is
\begin{equation}
 u_w=\frac{i\beta}{2}\exp\left(\frac{k\pi}{2\gamma\epsilon}\right)
 \exp\left(-iq+i\delta_w\right)\left[
 \left(1+5\gamma^2\epsilon^2\right)Q_0(q)+\gamma^4\epsilon^4Q_2(q)
 \ldots\right] \,, \label{equwqepsexp}
\end{equation}
where
\begin{align}
 Q_0(q)&=1+\frac{6i}{q}-\frac{33}{q^2}-\frac{237i}{q^3}
 +\frac{1890}{q^4}+\frac{17028i}{q^5}
 -\frac{167733}{q^6}+\ldots \,, \label{eqq0q} \\
 Q_2(q)&=-\frac{2iq^3}{15}-\frac{q^2}{5}
 +\frac{39iq}{5}-25+\frac{234i}{q}
 -\frac{14343}{5q^2}-\frac{181119i}{5q^3}+\ldots \,. \label{eqq2q}
\end{align}
We should like to point out that Eq.\ \eqref{equwqepsexp} disagrees with Eq.\ $(48)$ of Ref.\ \cite{Grimshaw95}, where the factor $(1+5\gamma^2\epsilon^2)$ is missing.
The truncated versions of the asymptotic expansions $Q_n(q)$ correspond to complex extensions of symmetric functions. Hence, for $\delta_w=0$ the function $u_w$ is purely imaginary on the imaginary $q$ axis, corresponding to the complex extension of an antisymmetric function. If $\delta_w=\pi/2$ then $u_w$ is real on the imaginary axis, and it corresponds to a symmetric function. Remember that $\beta$ and $\delta_w$ may also have $\epsilon$ dependence.

\subsection{Amplitude up to third order}

There are two important solutions of Eq.~\eqref{equxxxx}, the symmetric solution $u_m$ that has minimal tail, and the asymmetric solution $u_{-}$ that has no tail for $x>0$. They both can be calculated by numerical methods, but no accurate analytical solutions are known for either of them. However, their difference can be determined very precisely by the earlier presented higher order WKB method. The difference $u_m-u_{-}$ is exponentially small in terms of $\epsilon$, not just in the tail but also in the core domain, and even on the complex $x$ plane, including the matching region near the singularity. This allows us to represent the difference with the linearized solution $u_w$ in \eqref{equwnoexp} with appropriate tail-amplitude $\beta$ and phase $\delta_w$. The symmetric solution $u_m$ has the oscillating tail given by \eqref{equmasympt}, with amplitude $\alpha_m$ and phase $\delta_m$. The tail for $x>0$ can be compensated by $u_w$ to obtain the asymmetric solution,
\begin{equation}
 u_{-}=u_m-u_w \quad \mathrm{for} \quad \beta=\alpha_m \,,
 \ \ \delta_w=0 \,. \label{equminumuw}
\end{equation}
This equation is also valid on the complex $q$ plane where $u_w$ is given by \eqref{equwqepsexp}.

Although the difference $u_w$ is relatively small with respect to $u_m$ in the core region and on its complex extension, there is a place where it can be clearly observed. Because of its symmetry, $u_m$ is purely real on the imaginary axis, on both the complex $x$ and $q$ planes. Hence it follows from \eqref{equminumuw} that the imaginary parts of $u_{-}$ and $-u_w$ have to agree there,
\begin{equation}
 \mathrm{Im}\,u_{-}=-\mathrm{Im}\,u_w \quad
 \mathrm{for} \ \  \mathrm{Re}\,q=0 \,, \
 \mathrm{Im}\,q<0 \,, \label{eqimumeqm}
\end{equation}
where $\beta=\alpha_m$ and $\delta_w=0$ in the form \eqref{equwqepsexp} of $u_w$. This equality is valid to any order in $\epsilon$ and $1/q$. The solution of the inner problem discussed earlier can be used to obtain the imaginary part of $u_{-}$ on the imaginary axis. The importance of \eqref{eqimumeqm} is that after this we can directly obtain the minimal tail amplitude $\alpha_m=\beta$ as well.

To leading order in $\epsilon$ we can use \eqref{eqimv0m} to determine the imaginary part of $u_{-}$ on the imaginary axis. Using $v=\epsilon^2 u$ in \eqref{eqvepsepsu} and $v\approx v_0$ in \eqref{eqvvnepsnn} we obtain that
\begin{equation}
 \mathrm{Im}\,u_{-}=\frac{\pi K}{2\epsilon^2} \exp(-iq)
 \left(1+O(\epsilon^2)\right) \quad
 \mathrm{for} \ \  \mathrm{Re}\,q=0 \,, \
 \mathrm{Im}\,q<0 \,. \label{eqimumpikk1}
\end{equation}
Comparing this with $\mathrm{Im}\,u_w$ according to \eqref{eqimumeqm}, considering the leading order part of \eqref{equwqepsexp}, since $\alpha_m=\beta$, we obtain the expression for the minimal tail amplitude,
\begin{equation}
 \alpha_m^{(k,1)}=-\frac{\pi K}{\epsilon^2}
 \exp\left(-\frac{k\pi}{2\gamma\epsilon}\right)
  \,. \label{eqalpmk1}
\end{equation}
The notation $k$ in the upper index indicates that according to \eqref{eqksqrt} we keep $k=\sqrt{1+4\gamma^2\epsilon^2}$ inside the exponential. Writing $1$ in the upper index shows that this result is precise to linear order in $\epsilon$, since the expression is valid up to a factor of $(1+O(\epsilon^2))$.

\begin{figure}[!hbt]
 \centering
 \includegraphics[width=115mm]{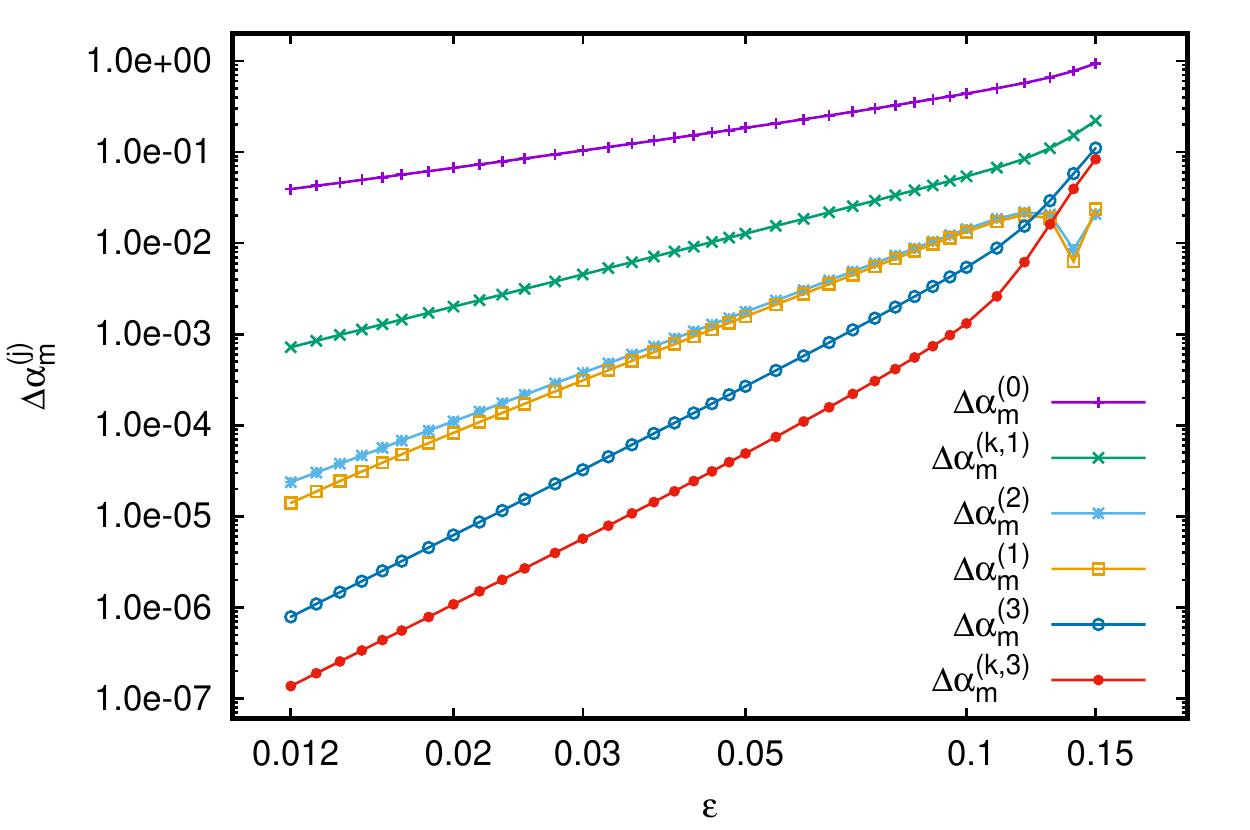}
\caption{\label{figaldiff3} Log-log plot of $\Delta\alpha_m^{(j)}=|(\alpha_m-\alpha_m^{(j)})/\alpha_m|$, showing the relative difference of the numerically calculated $\alpha_m$ from its various order analytic approximations $\alpha_m^{(j)}$ up to order three in $\epsilon$.}
\end{figure}

To leading order in $\epsilon$ we can substitute $k=1$, and we obtain the result first derived by Pomeau \textit{et al.}~in \cite{Pomeau88},
\begin{equation}
 \alpha_m^{(0)}=-\frac{\pi K}{\epsilon^2}
 \exp\left(-\frac{\pi}{2\gamma\epsilon}\right)\,. \label{eqalpm0}
\end{equation}
In Eq.~$(21)$ of \cite{Pomeau88} the double tail amplitude has been calculated for the asymmetric solution which tends to zero exponentially for $x\to-\infty$. Furthermore, there is an additional unnecessary factor of $2$ there because of a $1/2$ lost earlier, and an obviously missing fraction slash in the exponential. In Fig.~\ref{figaldiff3} we compare the various order analytic results $\alpha_m^{(j)}$ in this section to the numerical amplitude $\alpha_m$ calculated by the high precision spectral method.
We plot logarithmically the relative difference $\Delta\alpha_m^{(j)}=|(\alpha_m-\alpha_m^{(j)})/\alpha_m|$ as a function of $\epsilon$. Since the numerical result is much more accurate in these cases, $\Delta\alpha_m^{(j)}$ shows the relative error of the $j$th order analytic expansion result.

With the $k$ factor included in the exponential, the result $\alpha_m^{(k,1)}$ for the tail amplitude in \eqref{eqalpmk1} corresponds to Eq.~$(56)$ of Grimshaw and Joshi \cite{Grimshaw95}.
Since
\begin{align}
 \exp\left(-\frac{k\pi}{2\gamma\epsilon}\right)=
 \exp\left(-\frac{\pi}{2\gamma\epsilon}\right)
 &\left[ 1-\pi\gamma\epsilon+\frac{\pi^2}{2}\gamma^2\epsilon^2
 -\left(\frac{\pi^2}{6}-1\right)\pi\gamma^3\epsilon^3
 +\left(\frac{\pi^2}{24}-1\right)\pi^2\gamma^4\epsilon^4\right. \notag\\
 &\left.\ \
 -\left(\frac{\pi^4}{120}-\frac{\pi^2}{2}+2\right)\pi\gamma^5\epsilon^5
 +O(\epsilon^6)\right] \,, \label{eqexpkexp}
\end{align}
we can see that $\alpha_m^{(k,1)}$ contains odd powers of $\epsilon$ in its expansion. This shows that the inclusion of $k$ in the exponential improves the result to make it valid to linear $\epsilon$ order. Since \eqref{eqvvnepsnn} and \eqref{equwqepsexp} contain only even powers of $\epsilon$, odd powers in the expansion of $\alpha_m$ can only come from $k$ in the exponential. However, we will show below that contrary to the claim in \cite{Grimshaw95}, the result \eqref{eqalpmk1} is not valid to $\epsilon^2$ order.

Substituting the expansion \eqref{eqexpkexp} into \eqref{eqalpmk1} we can obtain an alternative first-order result for the amplitude,
\begin{equation}
 \alpha_m^{(1)}=-\frac{\pi K}{\epsilon^2}
 \exp\left(-\frac{\pi}{2\gamma\epsilon}\right)
 \left(1-\pi\gamma\epsilon\right)\,. \label{eqalpm1}
\end{equation}
Surprisingly, according to Fig.~\ref{figaldiff3}, $\alpha_m^{(1)}$ has much lower relative error than the other first order result $\alpha_m^{(k,1)}$. The reason for this is that the $\epsilon$ expansion of $\alpha_m^{(k,1)}$ contains an $\epsilon^2$ term with coefficient $\pi^2\gamma^2/2\approx 4.93\gamma^2$ which is much larger than the correct coefficient, which turns out to be $\pi^2\gamma^2/2-5\gamma^2\approx -0.07\gamma^2$ as we will see a bit later.

As we have already seen, Eq.~\eqref{eqv1qqqq} for the second order inner problem has an appropriate exact solution, $v_1=2/3$. This has no imaginary part on the imaginary $q$ axis. To this approximation \eqref{eqvvnepsnn} gives $v\approx v_0+\gamma^2\epsilon^2v_1$, hence \eqref{eqimumpikk1} is also valid to $\epsilon^2$ order,
\begin{equation}
 \mathrm{Im}\,u_{-}=\frac{\pi K}{2\epsilon^2} \exp(-iq)
 \left(1+O(\epsilon^4)\right) \quad
 \mathrm{for} \ \  \mathrm{Re}\,q=0 \,, \
 \mathrm{Im}\,q<0 \,. \label{eqimumpikk2}
\end{equation}
On the other hand, Eq.~\eqref{equwqepsexp} clearly has an $\epsilon^2$ part, because of the $\left(1+5\gamma^2\epsilon^2\right)$ factor. In order to make \eqref{eqimumeqm} valid to order $\epsilon^2$ we have to cancel this contribution by an $\epsilon$ dependent factor in the $\beta=\alpha_m$ amplitude. This way we obtain a higher order generalization of \eqref{eqalpmk1} for the minimal amplitude,
\begin{equation}
 \alpha_m^{(k,3)}=-\frac{\pi K}{\epsilon^2}
 \exp\left(-\frac{k\pi}{2\gamma\epsilon}\right)
 \left(1-5\gamma^2\epsilon^2\right) \,. \label{eqalpmk3}
\end{equation}
This approximation is correct to $\epsilon^3$ order, since the next correction would be an $\epsilon^4$ term in the factor multiplying the exponential term. As we can see in Fig.~\ref{figaldiff3}, the result $\alpha_m^{(k,3)}$ is significantly more precise than the lower order approximations. For the lowest $\epsilon$ values considered, it gives the amplitude to seven digits of precision. The amplitude $\alpha_m^{(k,3)}$ is the corrected version of the result, Eq.~(56), in Ref.~\cite{Grimshaw95}. That expression is valid only to $\epsilon^1$ order. The correctness of this additional $5\gamma^2\epsilon^2$ term in \eqref{eqalpmk3} is clearly supported by the numerical simulations. Actually, the reported differences between the analytical amplitude in \cite{Grimshaw95} and the numerical results of Boyd in \cite{Boyd95} are due to this missing term.

Using the expansion \eqref{eqexpkexp} we can obtain the following second and third order results for the amplitude
\begin{align}
 \alpha_m^{(2)}&=-\frac{\pi K}{\epsilon^2}
 \exp\left(-\frac{\pi}{2\gamma\epsilon}\right)
 \left[1-\pi\gamma\epsilon
 +\left(\frac{\pi^2}{2}-5\right)\gamma^2\epsilon^2\right] \,, \\
 \alpha_m^{(3)}&=-\frac{\pi K}{\epsilon^2}
 \exp\left(-\frac{\pi}{2\gamma\epsilon}\right)
 \left[1-\pi\gamma\epsilon
 +\left(\frac{\pi^2}{2}-5\right)\gamma^2\epsilon^2
 -\left(\frac{\pi^2}{6}-6\right)\pi\gamma^3\epsilon^3
 \right] \,.
\end{align}
Strangely, as we can see in Fig.~\ref{figaldiff3}, $\alpha_m^{(2)}$ gives slightly less precise approximation than the lower order result $\alpha_m^{(1)}$ given earlier in \eqref{eqalpm1}. The reason for this is that for the presented range of $\epsilon$ the neglected third order term is still larger than the second order one. Their orders become reversed only below $\epsilon\approx 0.00477$, which is not accessible by our numerical code. However, when the $\epsilon^3$ term is included, we obtain $\alpha_m^{(3)}$, which, according to the figure, is almost as precise as $\alpha_m^{(k,3)}$ in \eqref{eqalpmk3}. As can be expected, all $\Delta\alpha_m^{(j)}$ and $\Delta\alpha_m^{(k,j)}$ tend to zero as $\epsilon^{j+1}$ in the figure, except $\Delta\alpha_m^{(1)}$ which decays faster in the presented $\epsilon$ interval. In the next section we will determine the $\epsilon^4$ order correction to the amplitude in \eqref{eqalpmk3}. The fact that we can obtain even higher order results gives a very strong support that the results in the present section are correct.

The minimal amplitude tail has the asymptotic phase $\delta_m$. As we have already seen, the tail amplitude $\alpha$ of general symmetric solutions with asymptotic tail phase $\delta$ can be calculated from the minimal amplitude $\alpha_m$ by \eqref{eqalpsalpm}.

\section{Fourth order} \label{secfourth}

\subsection{Inner problem}

In this case the equations we have to solve follow from the substitution of the expansion \eqref{eqvvnepsnn} of $v$ into the inner Eq.~\eqref{eqvinnerqqqq}. The fist two equations have already been given in \eqref{eqv0qqqq} and \eqref{eqv1qqqq}. After substituting $v_1=2/3$ the next equation is
\begin{equation}
 v_{2\,qqqq}+v_{2\,qq}+6v_0 v_2-16v_0
 -\frac{4}{3} = 0 \,. \label{eqv2qqqq}
\end{equation}
The function $v_0$ is assumed to be known here. Although we do not know an exact solution for $v_0$, we can use the power series asymptotic expansion \eqref{eqv0expq} which starts according to \eqref{eqv0qexp}. There may also be an exponentially small correction, corresponding to \eqref{eqv0mmv0s}, and given in more detail in \eqref{eqw0qexp} of Appendix \ref{applininner}.

Since \eqref{eqv2qexp} shows that the expansion of $v_2$ consists only of even powers of $q$, and starts with a $q^2$ term, we look for solutions in the form
\begin{equation}
 v_2=\sum_{j=-1}^{\infty}b^{(2)}_j q^{-2j} \,. \label{eqvkexpq}
\end{equation}
Substituting into \eqref{eqv2qqqq} we obtain that $b^{(2)}_{-1}=-2/15$, and for $n\geq 0$ we get
\begin{equation}
 (2n)(2n+1)b^{(2)}_n+(2n-2)(2n-1)(2n)(2n+1)b^{(2)}_{n-1}
 +6\sum_{j=-1}^{n}b^{(2)}_{j}b^{(0)}_{n-j+1}
 -16 b^{(0)}_{n+1}=0 \,. \label{eqbn2gen}
\end{equation}
Separating the last term from the summation we can obtain a recursion relation, similar to \eqref{eqbn0rec}. The coefficients $b^{(2)}_{n}$ that we get agree with those already given in \eqref{eqv2qexp}. This shows that we match the inner problem appropriately to the outer problem.

\subsection{Laplace transform}

We intend to apply the Laplace transform method that we have already used for the zeroth order in \eqref{eqv0int}. Our aim is to calculate the imaginary part of the asymmetric function $v_2^{(-)}$ on the imaginary $q$ axis. However, for the $j\geq 0$ integer we cannot obtain $q^j$ as the Laplace transform of a smooth function. Hence we have to separate those terms, defining $\tilde v_2$ by
\begin{equation}
 v_2=-\frac{2}{15}q^2+\frac{2}{3}+\tilde v_2 \,. \label{eqv2tildev2}
\end{equation}
We can now look for $\tilde v_2$ in the form of the Laplace transform of a function $V'_2(s)$,
\begin{equation}
 \tilde v_2=\int_\Gamma I_2(s)\mathrm{d}s \,,\quad \quad
 I_2(s)=\exp(-sq)V'_2(s)
 \,, \label{eqtilvkint}
\end{equation}
where the contour $\Gamma$ is from $s=0$ to infinity, satisfying $\mathrm{Re}(sq)>0$. Similar to \eqref{eqvvpr0} and \eqref{eqan0bnp10}, from the identity for the Laplace transform of powers of $s$ it follows that
\begin{equation}
 V'_2(s)=\sum_{n=0}^{\infty}a^{(2)}_n s^{2n+1} \,, \label{eqvvprk}
\end{equation}
where
\begin{equation}
 a^{(2)}_n=\frac{b^{(2)}_{n+1}}{(2n+1)!} \,. \label{eqan0bnpkn}
\end{equation}
Since $b^{(2)}_1=64$, it follows that $a^{(2)}_0=64$. Substituting into \eqref{eqbn2gen}, for $n\geq 1$ we get
\begin{align}
 a^{(2)}_n+a^{(2)}_{n-1}
 &+\frac{6}{(2n+3)!}\sum_{j=0}^{n}
 (2j+1)!(2n-2j+1)!\,a^{(2)}_{j}a^{(0)}_{n-j} \notag\\
 &-\frac{4}{5}(2n+5)(2n+4)a^{(0)}_{n+2}
 -12a^{(0)}_{n+1}=0 \,. \label{eqan2gen}
\end{align}
The recursion relation is
\begin{align}
 \frac{(n+3)(2n-1)}{(n+1)(2n+3)}a^{(2)}_n=&-a^{(2)}_{n-1}
 -\frac{6}{(2n+3)!}\sum_{j=0}^{n-1}
 (2j+1)!(2n-2j+1)!\,a^{(2)}_{j}a^{(0)}_{n-j} \notag\\
 &+\frac{4}{5}(2n+5)(2n+4)a^{(0)}_{n+2}
 +12a^{(0)}_{n+1} \,. \label{eqan2rec}
\end{align}
For large $n$, the leading order behavior is $a^{(2)}_n\sim n^3(-1)^n$. It follows that the series for $V'_2(s)$ is convergent. The singularities are again located at $s=\pm n i$, where $n$ is any positive integer, but not at $s=0$. To study the function near the $s=i$ singularity we need to study the large $n$ behavior of the coefficients $a^{(2)}_n$. We look for the large $n$ behavior in the form
\begin{equation}
 a^{(2)}_n=(-1)^n\,\tilde a^{(2)}_n \,,\quad \qquad
 \tilde a^{(2)}_n=\sum_{l=-3}^\infty\frac{g_l^{(2)}}{n^l}
 \,. \label{eqakngn}
\end{equation}
For $\tilde a^{(0)}_n$ we use the expansion \eqref{eqa0ngn} with the already calculated coefficients given in \eqref{eqa0nexpn}. We substitute these into the equation that follows from \eqref{eqan2gen},
\begin{align}
 \tilde a^{(2)}_n-\tilde a^{(2)}_{n-1}
 &+\sum_{j=0}^{j_m}(-1)^j\, W_{n,j}^{(0)}\,\tilde a^{(2)}_{n-j}
 +\sum_{j=0}^{j_m}(-1)^j\, W_{n,j}^{(2)}\,\tilde a^{(0)}_{n-j} \notag\\
 &-\frac{4}{5}(2n+5)(2n+4)\tilde a^{(0)}_{n+2}
 +12\tilde a^{(0)}_{n+1}=0\,, \label{eqan2til}
\end{align}
where $W_{n,j}^{(0)}$ is defined in \eqref{eqwnk0}, $j_m$ is the positive integer used there, and
\begin{equation}
 W_{n,j}^{(2)}=\frac{6}{(2n+3)!}(2j+1)!(2n-2j+1)!\,a^{(2)}_{j} \,.
\end{equation}
Equation \eqref{eqan2til} has a homogeneous part that can be obtained by setting $\tilde a^{(0)}_{n}=0$. Note that the value of $W_{n,j}^{(0)}$ is nonzero even in this case.
The homogeneous part agrees exactly with Eq.~\eqref{eqanm1ankm}; hence, it has the solution $G_0(n)$ multiplied by an arbitrary constant. It follows that the general solution of \eqref{eqan2til} can be written as
\begin{equation}
 \tilde a^{(2)}_{n}=K\left(G_2(n)+K_2 G_0(n)\right) \,, \label{eqatinj}
\end{equation}
where $K_2$ is an arbitrary constant and $G_2(n)$ is a particular solution. Its expansion can be obtained using some algebraic manipulation software, similar to \eqref{eqa0nexpn},
\begin{equation}
 G_2(n)=\frac{16}{15}n^3+\frac{28}{5}n^2
 +\frac{368}{15}n-\frac{132}{n}+\frac{4122}{5n^2}
 -\frac{50833}{10n^3}+\frac{3144915}{112n^4}+\ldots \,. \label{eqgg2n}
\end{equation}
We have chosen the unique particular solution $G_2(n)$ for which there is no $n^0$ term. We can obtain the value of the constant $K_2$ by calculating $a^{(2)}_{n}$ using \eqref{eqan2rec} for large $n$, similar to what we did for $K$ in \eqref{eqkkmanydig}. Using $K_2\approx[a^{(2)}_n(-1)^n/K-G_2(n)]/G_0(n)$ we get
\begin{equation}
 K_2=-36.544068193583744293  \,.
\end{equation}

\subsection{Asymmetric solution}

Our aim is to calculate the imaginary part of the asymmetric $v_2\equiv v_2^{(-)}$ solution on the lower part of the imaginary $q$ axis. Obviously, this will be the same as the imaginary part of $\tilde v_2\equiv\tilde v_2^{(-)}$, since they differ only by two terms according to \eqref{eqv2tildev2}. We apply the same reasoning as we did for $v_0^{(-)}$ in the paragraph before Eq.~\eqref{eqsumexpmsq}. The function $\tilde v_j^{(-)}$ can be calculated in the domain $-\pi<\arg(q)\leq 0$ by the integral \eqref{eqtilvkint}, with a contour in the region satisfying $0<\arg(s)<\pi/2$. The imaginary part of $\tilde v_2^{(-)}$ on the imaginary axis can be determined to leading order using the residue of the function $I_2(s)$ at $s=i$. Substituting \eqref{eqvvprk}, \eqref{eqakngn} and \eqref{eqatinj} we get
\begin{equation}
 I_2(s)\approx\exp(-sq)K\sum_{n=0}^\infty (-1)^n
 \left[G_2(n)+K_2 G_0(n)\right]s^{2n+1} \,.
\end{equation}
From the expansion of $G_0(n)$ and $G_2(n)$ only terms of the form $(-1)^n n^j s^{2n+1}$ with $j\geq 0$ will give contributions to this residue. These terms can be summed, for example for $j=1$,
\begin{equation}
 \sum_{n=0}^\infty (-1)^n n s^{2n+1}=
 -\frac{s^3}{(1+s^2)^2} \,.
\end{equation}
The sum can be calculated for other concrete $j\geq 0$ values, but we could not find a general formula valid for arbitrary $j$. However, the general residue turns out to be very simple,
\begin{equation}
 \underset{s=i}{\mathrm{Res}}\sum_{n=0}^\infty (-1)^n n^j s^{2n+1}=
 \frac{1}{2}(-1)^j \quad \mathrm{for} \quad j\geq 0 \,.
\end{equation}
Hence from \eqref{eqa0nexpn} and \eqref{eqgg2n} we get
\begin{equation}
 \underset{s=i}{\mathrm{Res}}\,I_2(s)
 =\exp(-iq)\frac{K}{2}\left(-\frac{16}{15}+\frac{28}{5}
 -\frac{368}{15}+K_2\right)
 =\exp(-iq)\frac{K}{2}\left(K_2-20\right) \,.
\end{equation}

Next we define a symmetric function $v_2^{(m)}$ by taking the path $\Gamma$ along the upper half of the imaginary axis. The difference of $v_2^{(-)}$ and $v_2^{(m)}$ can be calculated similar to Eq.~\eqref{eqv0mmv0s}. The residue theorem can be applied by choosing a curve going from $s=0$ to infinity in the domain $0<\arg(s)<\pi/2$ and coming back along the upper half of the imaginary $s$ axis. This way we have to take into account half of the residue at $s=i$, so we get
\begin{equation}
 v_2^{(-)}-v_2^{(m)}\approx\frac{1}{2}\pi i K
 \exp(-iq)\left(K_2-20\right) \,. \label{eqv2mmv2m}
\end{equation}
Here we write $\approx$ because we have not considered yet the change in $v_2$ due to the modification of $v_0$ from $v_0^{(-)}$ to $v_0^{(m)}$ in \eqref{eqv2qqqq}. This will provide a finite number of terms proportional to $q^n\exp(-iq)$ with $n>0$ integers. The coefficients of those terms are completely fixed already by the leading order constant $K$, which determines $v_0^{(-)}-v_0^{(m)}$ according to \eqref{eqv0mmv0s}. The detailed analysis of the perturbations of Eq.~\eqref{eqv2qqqq} is given in Appendix \ref{applininner}. The change in $v_2$ due to the modification of $v_0$ can be chosen in a way that it does not influence the terms $q^0\exp(-iq)$ calculated in \eqref{eqv2mmv2m}.

Analogously to \eqref{eqimv0m}, taking the imaginary part we obtain that on the lower part of the imaginary axis
\begin{equation}
 \mathrm{Im}\,v_2^{(-)}\approx\frac{1}{2}\pi K
 \exp(-iq)\left(K_2-20\right)  \,. \label{eqimv2m}
\end{equation}
Here and in the following few equations the approximate equation sign indicates that we only consider the $q^0\exp(-iq)$ parts of the expressions.
Using $v=\epsilon^2 u$ and \eqref{eqvvnepsnn} we obtain a higher order generalization of \eqref{eqimumpikk2},
\begin{equation}
 \mathrm{Im}\,u_{-}\approx\frac{\pi K}{2\epsilon^2} \exp(-iq)
 \left[1+\left(K_2-20\right)\gamma^4\epsilon^4
 +O(\epsilon^6)\right]  \label{eqimumpikk6}
\end{equation}
for $\mathrm{Re}\,q=0$, $\mathrm{Im}\,q<0$.

\subsection{Tail amplitude up to fifth order}

\begin{figure}[!hbt]
 \centering
 \includegraphics[width=115mm]{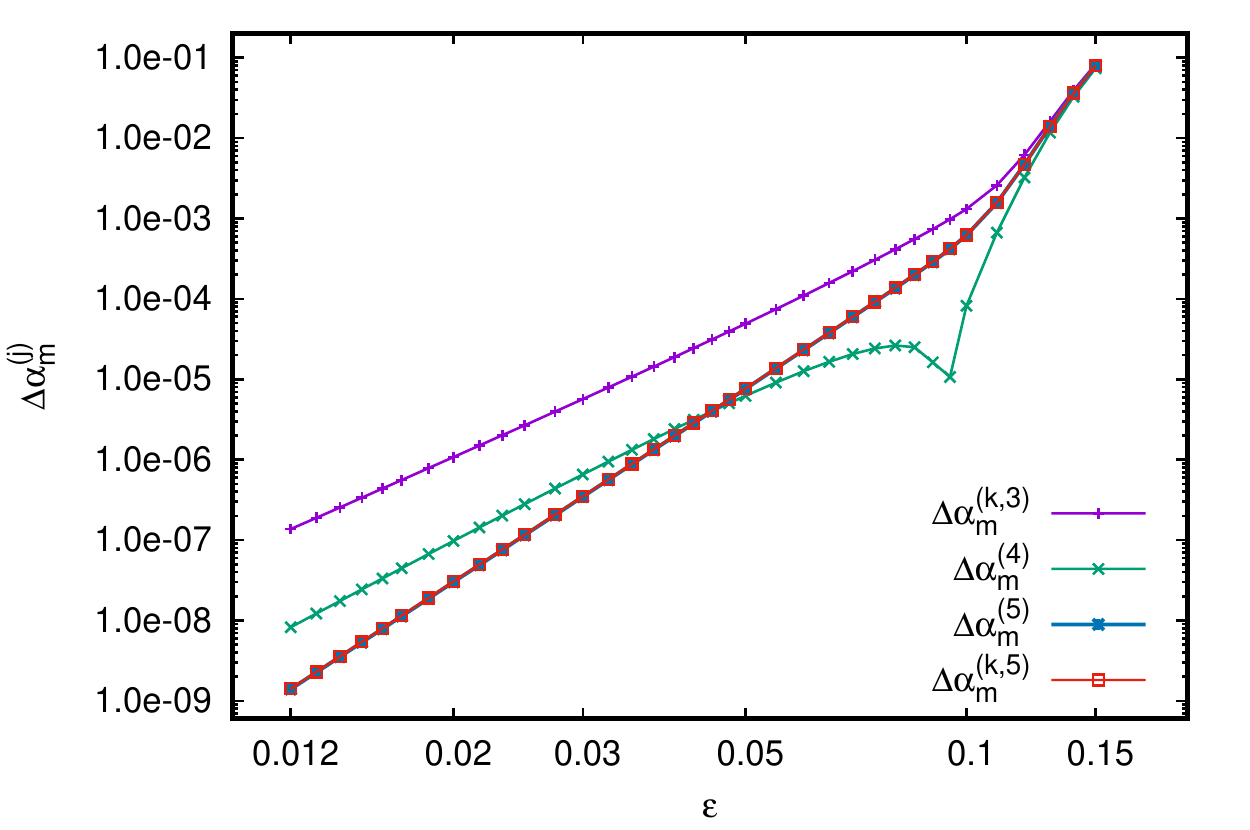}
\caption{\label{figaldiff5} Log-log plot of $\Delta\alpha_m^{(j)}=|(\alpha_m-\alpha_m^{(j)})/\alpha_m|$, showing the relative difference of the precise numerical $\alpha_m$ from various order analytic results $\alpha_m^{(j)}$, up to order five in $\epsilon$.}
\end{figure}

We generalize the expression for minimal amplitude $\alpha_m$ in \eqref{eqalpmk3} by allowing a fourth order contribution,
\begin{equation}
 \alpha_m^{(k,5)}=-\frac{\pi K}{\epsilon^2}
 \exp\left(-\frac{k\pi}{2\gamma\epsilon}\right)
 \left(1-5\gamma^2\epsilon^2-\xi_2\gamma^4\epsilon^4
 \right)\,, \label{eqalpmk5}
\end{equation}
where $\xi_2$ is a constant that will be determined below. This approximation is correct to $\epsilon^5$ order, since the next correction would be proportional to $\epsilon^6$.
Substituting $\beta=\alpha_m^{(k,5)}$ and $\delta_w=0$ into \eqref{equwqepsexp} and only considering the $q^0\exp(-iq)$ terms, we get
\begin{equation}
 u_w\approx-\frac{i\pi K}{2\epsilon^2} \exp(-iq)
 \left[1-\left(\xi_2+50\right)\gamma^4\epsilon^4
 +O(\epsilon^6) \right]\,.
\end{equation}
According to \eqref{eqimumeqm}, we compare the imaginary part on the axis with \eqref{eqimumpikk6} to obtain $K_2-20=-\xi_2-50$, which gives
\begin{equation}
 \xi_2=-K_2-30\approx 6.544068193583744293 \,.
\end{equation}
In Fig.~\ref{figaldiff5} we show the relative error of the analytic results $\alpha_m^{(j)}$ obtained in this section when compared to the precise numerical amplitude $\alpha_m$.
As can be expected, the relative error of the $j$th order results tend to zero as $\epsilon^{j+1}$.

Using the expansion \eqref{eqexpkexp} we can obtain the following alternative fifth order result for the amplitude,
\begin{align}
 \alpha_m^{(5)}=-\frac{\pi K}{\epsilon^2}
 \exp\left(-\frac{\pi}{2\gamma\epsilon}\right)
 &\left[1-\pi\gamma\epsilon
 +\left(\frac{\pi^2}{2}-5\right)\gamma^2\epsilon^2
 -\left(\frac{\pi^2}{6}-6\right)\pi\gamma^3\epsilon^3 \right. \label{res5}\\
 &\ \left.
 +\left(\frac{\pi^4}{24}-\frac{7\pi^2}{2}-\xi_2\right)\gamma^4\epsilon^4
 -\left(\frac{\pi^4}{120}-\frac{4\pi^2}{3}+7-\xi_2\right)
 \pi\gamma^5\epsilon^5
 \right]\,. \notag
\end{align}
We can also define a fourth order amplitude $\alpha_m^{(4)}$ by dropping the $\epsilon^5$ term from $\alpha_m^{(5)}$. The relative error of these approximations are also shown in Fig.~\ref{figaldiff5}. The two different fifth order results are so close to each other that $\Delta\alpha_m^{(5)}$ and $\Delta\alpha_m^{(k,5)}$ are indistinguishable in the logarithmic figure. Actually, $\Delta\alpha_m^{(k,5)}$ is larger by less than 2\% than $\Delta\alpha_m^{(5)}$. Both $\alpha_m^{(5)}$ and $\alpha_m^{(k,5)}$ are correct to nine digits of precision for the smallest $\epsilon$ value shown.

\section{Conclusions}\label{sec:conc}
We have considered the problem to compute corrections to the beyond-all-orders small amplitude standing wave tails of weakly localized soliton solutions in a fifth order KdV equation - the KdV eq.\ with an added fifth order dispersion term, $\epsilon^2\partial_x^5$.
The fKdV equation is not only of some genuine physical interest, but at the same time it also serves for us as a simplified model
to prepare the ground for the significantly more complicated problem for oscillon-quasibreathers in field theories.
For $\epsilon\ll1$ the simplest stationary solution is a bounded, one-parameter family of KdV-type solitons which are weakly localized due to an asymptotic standing wave tail, tending to the KdV 1-soliton for $\epsilon\to0$. These solutions are symmetric, and they are characterized by the asymptotic phase, $\delta$, of their tail.
Our main analytical result for the physically relevant \textsl{minimal amplitude} wave tail can be succinctly presented as
$$
 \alpha_{\rm min}=\frac{\lambda}{\epsilon^2}
 e^{-\frac{k(\epsilon)\pi}{2\gamma\epsilon}}
 \left(1-5\gamma^2\epsilon^2-\zeta_4\gamma^4\epsilon^4 +{\cal{O}}(\epsilon^6)
 \right) \,,\quad \lambda\approx19.9689\pi,\ \zeta_4\approx6.5441\,,
$$
where $\gamma$ parametrizes the speed of the unperturbed KdV soliton and the wave-number, $k(\epsilon)=(1+4\gamma^2\epsilon^2)^{1/2}$, has to be expanded up to ${\cal{O}}(\epsilon^5)$ [see Eq.\ \eqref{res5}].
Our paper resolves a long-standing discrepancy for the ${\cal{O}}(\epsilon^2)$ coefficient between the result of Ref.\ \cite{Grimshaw95} and the numerical results of Ref.\ \cite{Boyd95}.

The asymptotic phase of the minimal amplitude tail, $\delta_{\rm min}$, has been determined in a WKB approximation to rather high ($\sim 100$) orders in $\epsilon$.
We have also found that for a given $\epsilon$, the tail amplitude, $\alpha$, of any member of the one-parameter family, characterized with phase $\delta$,
is related to the minimal tail amplitude $\alpha_{\rm min}$ by
$ \alpha=\alpha_{\rm min}/\cos\left(\delta -\delta_{\rm min}\right)$.

We have also developed an efficient, arbitrary precision pseudospectral code to solve the fKdV equation, to investigate the asymptotic standing wave tails to high numerical precision.
Our numerical code is fast enough so that we can employ numerical minimization procedures to find the value of the phase, $\delta_{\rm min}$, of the minimal amplitude solution very precisely, even for quite small values of $\epsilon$. The numerically obtained value for $\delta_{\rm min}$ agrees very well with the optimally truncated asymptotic series result (see Fig.\ \ref{figdediff}).
The remarkably good agreement of our higher order perturbative results for $\alpha_{\rm min}$ with our numerical calculations (see Figs.\ \ref{figaldiff3} and \ref{figaldiff5}) gives very strong support that both our analytic and numerical considerations are correct and reliable.

\appendix

\section{Linear correction to the inner problem} \label{applininner}

The complex extension of the exponentially small linear correction $u_w$ to the outer solution has been calculated in Section \eqref{secexpsmall}. Similar small corrections can also be determined to $v_n$ at each order of the inner problem. We proceed by substituting $v_n\to v_n+w_n$ for all $n$ and linearizing for $w_n$.

From \eqref{eqv0qqqq} we obtain
\begin{equation}
 w_{0\,qqqq}+w_{0\,qq}+6v_0w_{0} = 0 \,. \label{eqw0lin}
\end{equation}
For $v_0$ we can use the asymptotic expansion \eqref{eqv0expq} that starts with the terms given in \eqref{eqv0qexp}. We search the solution for $w_0$ in the form
\begin{equation}
 w_0=\frac{i}{2}b_0\exp(-iq) F_0(q) \quad \,, \qquad
 F_0(q)=1+\sum_{j=1}^\infty \frac{w_{0}^{(j)}}{q^j} \,, \label{eqw0qexp}
\end{equation}
where $b_0$ and $w_{0}^{(j)}$ are complex constants. Substituting into \eqref{eqw0lin}, the coefficients of the various powers of $1/q$ determine the constants $w_{0}^{(j)}$. The first few terms give
\begin{equation}
 F_0(q)=
 1+\frac{6i}{q}-\frac{33}{q^2}-\frac{237i}{q^3}
 +\frac{1890}{q^4}+\frac{17028i}{q^5}
 -\frac{167733}{q^6}+\ldots \,. \label{eqf0q}
\end{equation}
This naturally agrees with the function $Q_0(q)$ given in \eqref{eqq0q}. Since Eq.~\eqref{eqw0lin} is linear, the constant $b_0$ can be arbitrary. Real $b_0$ corresponds to a function antisymmetric on the real axis, while purely imaginary $b_0$ to a symmetric one.

The linearization of the $\epsilon^2$ order inner equation \eqref{eqv1qqqq} gives
\begin{equation}
 w_{1\,qqqq}+w_{1\,qq}+6v_0 w_1+2(3v_1-2)w_0 = 0 \,.
\end{equation}
For $v_1=2/3$ this equation becomes the same as \eqref{eqw0lin}, and the solutions are
\begin{equation}
 w_1=\frac{i}{2}b_1\exp(-iq) F_0(q) \,, \label{eqw1qexp}
\end{equation}
where $b_1$ is arbitrary complex constant. Since $v_1=2/3$ satisfies the matching condition to the outer solution, and it is symmetric, without an imaginary part for $\mathrm{Re}\,q=0$, we do not need to perturb $v_1$, we can set $b_1=0$. On the other hand, the $\epsilon^2$ part of \eqref{equwqepsexp} contains a perturbation corresponding to \eqref{eqw1qexp} with $b_1=5b_0$. We had to cancel these terms by a $1-5\gamma^2\epsilon^2$ factor in the amplitude $\alpha_m$ in \eqref{eqalpmk3}.

Linearizing \eqref{eqv2qqqq} we obtain
\begin{equation}
 w_{2\,qqqq}+w_{2\,qq}+6v_0 w_2+2(3v_2-8)w_0 = 0 \,.
\end{equation}
The homogeneous part agrees with Eq.~\eqref{eqw0lin} for $w_0$; hence we can always add an arbitrary constant times $\exp(-iq)F_0(q)$ to the solution. Searching the solution in the form
\begin{equation}
 w_2=\frac{i}{2}\exp(-iq) \sum_{j=-3}^\infty \frac{w_{2}^{(j)}}{q^j} \,,
\end{equation}
from the coefficients of the various powers of $q$ we obtain
\begin{equation}
 w_2=\frac{i}{2}\exp(-iq)\left[b_0F_2(q)+b_2 F_0(q)\right] \,, \label{eqw2i2}
\end{equation}
where $b_0$ is the constant in \eqref{eqw0qexp}, $b_2$ is an arbitrary complex constant, the function $F_0(q)$ is given in \eqref{eqf0q}, and
\begin{equation}
 F_2(q)=-\frac{2i}{15}q^3-\frac{1}{5}q^2+\frac{39i}{5}q+\frac{384i}{q}
 -\frac{18468}{5q^2}-\frac{210744i}{5q^3}+\ldots \,.
\end{equation}
The function $F_2(q)$ was made unique here by setting the coefficient of the $q^0$ term zero. The $\epsilon^4$ part of \eqref{equwqepsexp}, which is proportional to the $Q_2(q)$ given in \eqref{eqq2q}, can be obtained from \eqref{eqw2i2} by setting $b_2=-25b_0$.

\section*{Acknowledgements}

The research of G.~F.~has been supported in part by the National Research Development and Innovation Office (NKFIH) OTKA Grants No.~K 138277 and No.~K 142423.

\bibliography{fkdv}

\begin{thebibliography}{30}%
\makeatletter
\providecommand \@ifxundefined [1]{%
 \@ifx{#1\undefined}
}%
\providecommand \@ifnum [1]{%
 \ifnum #1\expandafter \@firstoftwo
 \else \expandafter \@secondoftwo
 \fi
}%
\providecommand \@ifx [1]{%
 \ifx #1\expandafter \@firstoftwo
 \else \expandafter \@secondoftwo
 \fi
}%
\providecommand \natexlab [1]{#1}%
\providecommand \enquote  [1]{``#1''}%
\providecommand \bibnamefont  [1]{#1}%
\providecommand \bibfnamefont [1]{#1}%
\providecommand \citenamefont [1]{#1}%
\providecommand \href@noop [0]{\@secondoftwo}%
\providecommand \href [0]{\begingroup \@sanitize@url \@href}%
\providecommand \@href[1]{\@@startlink{#1}\@@href}%
\providecommand \@@href[1]{\endgroup#1\@@endlink}%
\providecommand \@sanitize@url [0]{\catcode `\\12\catcode `\$12\catcode
  `\&12\catcode `\#12\catcode `\^12\catcode `\_12\catcode `\%12\relax}%
\providecommand \@@startlink[1]{}%
\providecommand \@@endlink[0]{}%
\providecommand \url  [0]{\begingroup\@sanitize@url \@url }%
\providecommand \@url [1]{\endgroup\@href {#1}{\urlprefix }}%
\providecommand \urlprefix  [0]{URL }%
\providecommand \Eprint [0]{\href }%
\providecommand \doibase [0]{https://doi.org/}%
\providecommand \selectlanguage [0]{\@gobble}%
\providecommand \bibinfo  [0]{\@secondoftwo}%
\providecommand \bibfield  [0]{\@secondoftwo}%
\providecommand \translation [1]{[#1]}%
\providecommand \BibitemOpen [0]{}%
\providecommand \bibitemStop [0]{}%
\providecommand \bibitemNoStop [0]{.\EOS\space}%
\providecommand \EOS [0]{\spacefactor3000\relax}%
\providecommand \BibitemShut  [1]{\csname bibitem#1\endcsname}%
\let\auto@bib@innerbib\@empty
\bibitem [{\citenamefont {Copeland}\ \emph {et~al.}(1995)\citenamefont
  {Copeland}, \citenamefont {Gleiser},\ and\ \citenamefont
  {M{\"u}ller}}]{Copeland95}%
  \BibitemOpen
  \bibfield  {author} {\bibinfo {author} {\bibfnamefont {E.~J.}\ \bibnamefont
  {Copeland}}, \bibinfo {author} {\bibfnamefont {M.}~\bibnamefont {Gleiser}},\
  and\ \bibinfo {author} {\bibfnamefont {H.-R.}\ \bibnamefont {M{\"u}ller}},\
  }\bibfield  {title} {\bibinfo {title} {Oscillons: Resonant configurations
  during bubble collapse},\ }\href {https://doi.org/10.1103/PhysRevD.52.1920}
  {\bibfield  {journal} {\bibinfo  {journal} {Phys. Rev. D}\ }\textbf {\bibinfo
  {volume} {52}},\ \bibinfo {pages} {1920} (\bibinfo {year}
  {1995})}\BibitemShut {NoStop}%
\bibitem [{\citenamefont {Honda}\ and\ \citenamefont
  {Choptuik}(2002)}]{Honda2001}%
  \BibitemOpen
  \bibfield  {author} {\bibinfo {author} {\bibfnamefont {E.~P.}\ \bibnamefont
  {Honda}}\ and\ \bibinfo {author} {\bibfnamefont {M.~W.}\ \bibnamefont
  {Choptuik}},\ }\bibfield  {title} {\bibinfo {title} {Fine structure of
  oscillons in the spherically symmetric $\phi^4$ {Klein-Gordon} model},\
  }\href {https://doi.org/10.1103/PhysRevD.65.084037} {\bibfield  {journal}
  {\bibinfo  {journal} {Phys. Rev. D}\ }\textbf {\bibinfo {volume} {65}},\
  \bibinfo {pages} {084037} (\bibinfo {year} {2002})}\BibitemShut {NoStop}%
\bibitem [{\citenamefont {Cyncynates}\ and\ \citenamefont
  {Giurgica-Tiron}(2021)}]{Cyncynates2021}%
  \BibitemOpen
  \bibfield  {author} {\bibinfo {author} {\bibfnamefont {D.}~\bibnamefont
  {Cyncynates}}\ and\ \bibinfo {author} {\bibfnamefont {T.}~\bibnamefont
  {Giurgica-Tiron}},\ }\bibfield  {title} {\bibinfo {title} {Structure of the
  oscillon: The dynamics of attractive self-interaction},\ }\href
  {https://doi.org/10.1103/PhysRevD.103.116011} {\bibfield  {journal} {\bibinfo
   {journal} {Phys. Rev. D}\ }\textbf {\bibinfo {volume} {103}},\ \bibinfo
  {pages} {116011} (\bibinfo {year} {2021})}\BibitemShut {NoStop}%
\bibitem [{\citenamefont {Fodor}(2019)}]{Fodor19}%
  \BibitemOpen
  \bibfield  {author} {\bibinfo {author} {\bibfnamefont {G.}~\bibnamefont
  {Fodor}},\ }\bibfield  {title} {\bibinfo {title} {A review on radiation of
  oscillons and oscillatons},\ }\Eprint {https://arxiv.org/abs/1911.03340}
  {1911.03340}  (\bibinfo {year} {2019}),\ \bibinfo {note} {arXiv:1911.03340
  [hep-th]}\BibitemShut {NoStop}%
\bibitem [{\citenamefont {Visinelli}(2021)}]{Visinelli21}%
  \BibitemOpen
  \bibfield  {author} {\bibinfo {author} {\bibfnamefont {L.}~\bibnamefont
  {Visinelli}},\ }\bibfield  {title} {\bibinfo {title} {Boson stars and
  oscillatons: A review},\ }\href {https://doi.org/10.1142/S0218271821300068}
  {\bibfield  {journal} {\bibinfo  {journal} {International Journal of Modern
  Physics D}\ }\textbf {\bibinfo {volume} {30}},\ \bibinfo {pages} {2130006}
  (\bibinfo {year} {2021})}\BibitemShut {NoStop}%
\bibitem [{\citenamefont {Zhang}\ \emph {et~al.}(2020)\citenamefont {Zhang},
  \citenamefont {Amin}, \citenamefont {Copeland}, \citenamefont {Saffin},\ and\
  \citenamefont {Lozanov}}]{Zhang2020}%
  \BibitemOpen
  \bibfield  {author} {\bibinfo {author} {\bibfnamefont {H.-Y.}\ \bibnamefont
  {Zhang}}, \bibinfo {author} {\bibfnamefont {M.~A.}\ \bibnamefont {Amin}},
  \bibinfo {author} {\bibfnamefont {E.~J.}\ \bibnamefont {Copeland}}, \bibinfo
  {author} {\bibfnamefont {P.~M.}\ \bibnamefont {Saffin}},\ and\ \bibinfo
  {author} {\bibfnamefont {K.~D.}\ \bibnamefont {Lozanov}},\ }\bibfield
  {title} {\bibinfo {title} {Classical decay rates of oscillons},\ }\href
  {https://doi.org/10.1088/1475-7516/2020/07/055} {\bibfield  {journal}
  {\bibinfo  {journal} {Journal of Cosmology and Astroparticle Physics}\
  }\textbf {\bibinfo {volume} {2020}}\bibinfo  {number} { (07)},\ \bibinfo
  {pages} {055}}\BibitemShut {NoStop}%
\bibitem [{\citenamefont {Oll{\'e}}\ \emph {et~al.}(2021)\citenamefont
  {Oll{\'e}}, \citenamefont {Pujol{\`a}s},\ and\ \citenamefont
  {Rompineve}}]{Olle2020}%
  \BibitemOpen
\bibfield  {number} {  }\bibfield  {author} {\bibinfo {author} {\bibfnamefont
  {J.}~\bibnamefont {Oll{\'e}}}, \bibinfo {author} {\bibfnamefont
  {O.}~\bibnamefont {Pujol{\`a}s}},\ and\ \bibinfo {author} {\bibfnamefont
  {F.}~\bibnamefont {Rompineve}},\ }\bibfield  {title} {\bibinfo {title}
  {Recipes for oscillon longevity},\ }\href
  {https://doi.org/10.1088/1475-7516/2021/09/015} {\bibfield  {journal}
  {\bibinfo  {journal} {Journal of Cosmology and Astroparticle Physics}\
  }\textbf {\bibinfo {volume} {2021}}\bibinfo  {number} { (09)},\ \bibinfo
  {pages} {015}}\BibitemShut {NoStop}%
\bibitem [{\citenamefont {Fodor}\ \emph
  {et~al.}(2009{\natexlab{a}})\citenamefont {Fodor}, \citenamefont
  {Forg{\'a}cs}, \citenamefont {Horv{\'a}th},\ and\ \citenamefont
  {Mezei}}]{Fodor09a}%
  \BibitemOpen
\bibfield  {number} {  }\bibfield  {author} {\bibinfo {author} {\bibfnamefont
  {G.}~\bibnamefont {Fodor}}, \bibinfo {author} {\bibfnamefont
  {P.}~\bibnamefont {Forg{\'a}cs}}, \bibinfo {author} {\bibfnamefont
  {Z.}~\bibnamefont {Horv{\'a}th}},\ and\ \bibinfo {author} {\bibfnamefont
  {M.}~\bibnamefont {Mezei}},\ }\bibfield  {title} {\bibinfo {title}
  {Computation of the radiation amplitude of oscillons},\ }\href
  {https://doi.org/10.1103/PhysRevD.79.065002} {\bibfield  {journal} {\bibinfo
  {journal} {Phys. Rev. D}\ }\textbf {\bibinfo {volume} {79}},\ \bibinfo
  {pages} {065002} (\bibinfo {year} {2009}{\natexlab{a}})}\BibitemShut
  {NoStop}%
\bibitem [{\citenamefont {Fodor}\ \emph
  {et~al.}(2009{\natexlab{b}})\citenamefont {Fodor}, \citenamefont
  {Forg{\'a}cs}, \citenamefont {Horv{\'a}th},\ and\ \citenamefont
  {Mezei}}]{Fodor09b}%
  \BibitemOpen
  \bibfield  {author} {\bibinfo {author} {\bibfnamefont {G.}~\bibnamefont
  {Fodor}}, \bibinfo {author} {\bibfnamefont {P.}~\bibnamefont {Forg{\'a}cs}},
  \bibinfo {author} {\bibfnamefont {Z.}~\bibnamefont {Horv{\'a}th}},\ and\
  \bibinfo {author} {\bibfnamefont {M.}~\bibnamefont {Mezei}},\ }\bibfield
  {title} {\bibinfo {title} {Radiation of scalar oscillons in 2 and 3
  dimensions},\ }\href {https://doi.org/10.1016/j.physletb.2009.03.054}
  {\bibfield  {journal} {\bibinfo  {journal} {Physics Letters B}\ }\textbf
  {\bibinfo {volume} {674}},\ \bibinfo {pages} {319} (\bibinfo {year}
  {2009}{\natexlab{b}})}\BibitemShut {NoStop}%
\bibitem [{\citenamefont {Segur}\ and\ \citenamefont
  {Kruskal}(1987)}]{SegurKruskal87}%
  \BibitemOpen
  \bibfield  {author} {\bibinfo {author} {\bibfnamefont {H.}~\bibnamefont
  {Segur}}\ and\ \bibinfo {author} {\bibfnamefont {M.~D.}\ \bibnamefont
  {Kruskal}},\ }\bibfield  {title} {\bibinfo {title} {Nonexistence of
  small-amplitude breather solutions in $\phi^{4}$ theory},\ }\href
  {https://doi.org/10.1103/PhysRevLett.58.747} {\bibfield  {journal} {\bibinfo
  {journal} {Phys. Rev. Lett.}\ }\textbf {\bibinfo {volume} {58}},\ \bibinfo
  {pages} {747} (\bibinfo {year} {1987})}\BibitemShut {NoStop}%
\bibitem [{\citenamefont {Fodor}\ \emph {et~al.}(2006)\citenamefont {Fodor},
  \citenamefont {Forg{\'a}cs}, \citenamefont {Grandcl{\'e}ment},\ and\
  \citenamefont {R{\'a}cz}}]{Fodor06}%
  \BibitemOpen
  \bibfield  {author} {\bibinfo {author} {\bibfnamefont {G.}~\bibnamefont
  {Fodor}}, \bibinfo {author} {\bibfnamefont {P.}~\bibnamefont {Forg{\'a}cs}},
  \bibinfo {author} {\bibfnamefont {P.}~\bibnamefont {Grandcl{\'e}ment}},\ and\
  \bibinfo {author} {\bibfnamefont {I.}~\bibnamefont {R{\'a}cz}},\ }\bibfield
  {title} {\bibinfo {title} {Oscillons and quasibreathers in the
  ${\ensuremath{\phi}}^{4}$ {Klein-Gordon} model},\ }\href
  {https://doi.org/10.1103/PhysRevD.74.124003} {\bibfield  {journal} {\bibinfo
  {journal} {Phys. Rev. D}\ }\textbf {\bibinfo {volume} {74}},\ \bibinfo
  {pages} {124003} (\bibinfo {year} {2006})}\BibitemShut {NoStop}%
\bibitem [{\citenamefont {Saffin}\ and\ \citenamefont
  {Tranberg}(2007)}]{Saffin07}%
  \BibitemOpen
  \bibfield  {author} {\bibinfo {author} {\bibfnamefont {P.~M.}\ \bibnamefont
  {Saffin}}\ and\ \bibinfo {author} {\bibfnamefont {A.}~\bibnamefont
  {Tranberg}},\ }\bibfield  {title} {\bibinfo {title} {Oscillons and
  quasi-breathers in {D}+1 dimensions},\ }\href
  {http://stacks.iop.org/1126-6708/2007/i=01/a=030} {\bibfield  {journal}
  {\bibinfo  {journal} {Journal of High Energy Physics}\ }\textbf {\bibinfo
  {volume} {2007}},\ \bibinfo {pages} {030} (\bibinfo {year}
  {2007})}\BibitemShut {NoStop}%
\bibitem [{\citenamefont {Kakutani}\ and\ \citenamefont
  {Ono}(1969)}]{Kakutani69}%
  \BibitemOpen
  \bibfield  {author} {\bibinfo {author} {\bibfnamefont {T.}~\bibnamefont
  {Kakutani}}\ and\ \bibinfo {author} {\bibfnamefont {H.}~\bibnamefont {Ono}},\
  }\bibfield  {title} {\bibinfo {title} {Weak non-linear hydromagnetic waves in
  a cold collision-free plasma},\ }\href {https://doi.org/10.1143/JPSJ.26.1305}
  {\bibfield  {journal} {\bibinfo  {journal} {Journal of the Physical Society
  of Japan}\ }\textbf {\bibinfo {volume} {26}},\ \bibinfo {pages} {1305}
  (\bibinfo {year} {1969})}\BibitemShut {NoStop}%
\bibitem [{\citenamefont {Kawahara}(1972)}]{Kawahara72}%
  \BibitemOpen
  \bibfield  {author} {\bibinfo {author} {\bibfnamefont {T.}~\bibnamefont
  {Kawahara}},\ }\bibfield  {title} {\bibinfo {title} {Oscillatory solitary
  waves in dispersive media},\ }\href {https://doi.org/10.1143/JPSJ.33.260}
  {\bibfield  {journal} {\bibinfo  {journal} {Journal of the Physical Society
  of Japan}\ }\textbf {\bibinfo {volume} {33}},\ \bibinfo {pages} {260}
  (\bibinfo {year} {1972})}\BibitemShut {NoStop}%
\bibitem [{\citenamefont {Hunter}\ and\ \citenamefont
  {Scheurle}(1988)}]{Hunter88}%
  \BibitemOpen
  \bibfield  {author} {\bibinfo {author} {\bibfnamefont {J.~K.}\ \bibnamefont
  {Hunter}}\ and\ \bibinfo {author} {\bibfnamefont {J.}~\bibnamefont
  {Scheurle}},\ }\bibfield  {title} {\bibinfo {title} {Existence of perturbed
  solitary wave solutions to a model equation for water waves},\ }\href
  {https://doi.org/10.1016/0167-2789(88)90054-1} {\bibfield  {journal}
  {\bibinfo  {journal} {Physica D: Nonlinear Phenomena}\ }\textbf {\bibinfo
  {volume} {32}},\ \bibinfo {pages} {253} (\bibinfo {year} {1988})}\BibitemShut
  {NoStop}%
\bibitem [{\citenamefont {Benilov}\ \emph {et~al.}(1993)\citenamefont
  {Benilov}, \citenamefont {Grimshaw},\ and\ \citenamefont
  {Kuznetsova}}]{Benilov93}%
  \BibitemOpen
  \bibfield  {author} {\bibinfo {author} {\bibfnamefont {E.}~\bibnamefont
  {Benilov}}, \bibinfo {author} {\bibfnamefont {R.}~\bibnamefont {Grimshaw}},\
  and\ \bibinfo {author} {\bibfnamefont {E.}~\bibnamefont {Kuznetsova}},\
  }\bibfield  {title} {\bibinfo {title} {The generation of radiating waves in a
  singularly-perturbed {Korteweg-de Vries} equation},\ }\href
  {https://doi.org/10.1016/0167-2789(93)90091-E} {\bibfield  {journal}
  {\bibinfo  {journal} {Physica D: Nonlinear Phenomena}\ }\textbf {\bibinfo
  {volume} {69}},\ \bibinfo {pages} {270} (\bibinfo {year} {1993})}\BibitemShut
  {NoStop}%
\bibitem [{\citenamefont {Amick}\ and\ \citenamefont {McLeod}(1991)}]{Amick91}%
  \BibitemOpen
  \bibfield  {author} {\bibinfo {author} {\bibfnamefont {C.~J.}\ \bibnamefont
  {Amick}}\ and\ \bibinfo {author} {\bibfnamefont {J.~B.}\ \bibnamefont
  {McLeod}},\ }\bibfield  {title} {\bibinfo {title} {A singular perturbation
  problem in water-waves},\ }\href@noop {} {\bibfield  {journal} {\bibinfo
  {journal} {Stability and Applied Analysis of Continuous Media}\ }\textbf
  {\bibinfo {volume} {1}},\ \bibinfo {pages} {127} (\bibinfo {year}
  {1991})}\BibitemShut {NoStop}%
\bibitem [{\citenamefont {Gunney}\ \emph {et~al.}(1999)\citenamefont {Gunney},
  \citenamefont {Li},\ and\ \citenamefont {Olver}}]{Gunney99}%
  \BibitemOpen
  \bibfield  {author} {\bibinfo {author} {\bibfnamefont {B.~T.~N.}\
  \bibnamefont {Gunney}}, \bibinfo {author} {\bibfnamefont {Y.~A.}\
  \bibnamefont {Li}},\ and\ \bibinfo {author} {\bibfnamefont {P.~J.}\
  \bibnamefont {Olver}},\ }\bibfield  {title} {\bibinfo {title} {Solitary waves
  in the critical surface-tension model},\ }\href
  {https://doi.org/10.1023/A:1004593923334} {\bibfield  {journal} {\bibinfo
  {journal} {Journal of Engineering Mathematics}\ }\textbf {\bibinfo {volume}
  {36}},\ \bibinfo {pages} {99} (\bibinfo {year} {1999})}\BibitemShut {NoStop}%
\bibitem [{\citenamefont {Boyd}(1990)}]{Boyd90}%
  \BibitemOpen
  \bibfield  {author} {\bibinfo {author} {\bibfnamefont {J.~P.}\ \bibnamefont
  {Boyd}},\ }\bibfield  {title} {\bibinfo {title} {A numerical calculation of a
  weakly non-local solitary wave: the $\phi^4$ breather},\ }\href
  {https://iopscience.iop.org/article/10.1088/0951-7715/3/1/010} {\bibfield
  {journal} {\bibinfo  {journal} {Nonlinearity}\ }\textbf {\bibinfo {volume}
  {3}},\ \bibinfo {pages} {177} (\bibinfo {year} {1990})}\BibitemShut {NoStop}%
\bibitem [{\citenamefont {Boyd}(1998)}]{Boyd98}%
  \BibitemOpen
  \bibfield  {author} {\bibinfo {author} {\bibfnamefont {J.~P.}\ \bibnamefont
  {Boyd}},\ }\href {https://doi.org/10.1007/978-1-4615-5825-5} {\emph {\bibinfo
  {title} {Weakly Nonlocal Solitary Waves and Beyond-All-Orders Asymptotics:
  Generalized Solitons and Hyperasymptotic Perturbation Theory}}}\ (\bibinfo
  {publisher} {Springer US},\ \bibinfo {address} {Boston, MA},\ \bibinfo {year}
  {1998})\BibitemShut {NoStop}%
\bibitem [{\citenamefont {Pomeau}\ \emph {et~al.}(1988)\citenamefont {Pomeau},
  \citenamefont {Ramani},\ and\ \citenamefont {Grammaticos}}]{Pomeau88}%
  \BibitemOpen
  \bibfield  {author} {\bibinfo {author} {\bibfnamefont {Y.}~\bibnamefont
  {Pomeau}}, \bibinfo {author} {\bibfnamefont {A.}~\bibnamefont {Ramani}},\
  and\ \bibinfo {author} {\bibfnamefont {B.}~\bibnamefont {Grammaticos}},\
  }\bibfield  {title} {\bibinfo {title} {Structural stability of the
  {Korteweg-de Vries} solitons under a singular perturbation},\ }\href
  {https://doi.org/10.1016/0167-2789(88)90018-8} {\bibfield  {journal}
  {\bibinfo  {journal} {Physica D: Nonlinear Phenomena}\ }\textbf {\bibinfo
  {volume} {31}},\ \bibinfo {pages} {127} (\bibinfo {year} {1988})}\BibitemShut
  {NoStop}%
\bibitem [{\citenamefont {Grimshaw}\ and\ \citenamefont
  {Joshi}(1995)}]{Grimshaw95}%
  \BibitemOpen
  \bibfield  {author} {\bibinfo {author} {\bibfnamefont {R.}~\bibnamefont
  {Grimshaw}}\ and\ \bibinfo {author} {\bibfnamefont {N.}~\bibnamefont
  {Joshi}},\ }\bibfield  {title} {\bibinfo {title} {Weakly nonlocal solitary
  waves in a singularly perturbed {Korteweg--De Vries} equation},\ }\href
  {https://doi.org/10.1137/S0036139993243825} {\bibfield  {journal} {\bibinfo
  {journal} {SIAM Journal on Applied Mathematics}\ }\textbf {\bibinfo {volume}
  {55}},\ \bibinfo {pages} {124} (\bibinfo {year} {1995})}\BibitemShut
  {NoStop}%
\bibitem [{\citenamefont {Boyd}(1991)}]{Boyd91}%
  \BibitemOpen
  \bibfield  {author} {\bibinfo {author} {\bibfnamefont {J.~P.}\ \bibnamefont
  {Boyd}},\ }\bibfield  {title} {\bibinfo {title} {Weakly non-local solitons
  for capillary-gravity waves: Fifth-degree {Korteweg-de Vries} equation},\
  }\href {https://doi.org/10.1016/0167-2789(91)90056-F} {\bibfield  {journal}
  {\bibinfo  {journal} {Physica D: Nonlinear Phenomena}\ }\textbf {\bibinfo
  {volume} {48}},\ \bibinfo {pages} {129} (\bibinfo {year} {1991})}\BibitemShut
  {NoStop}%
\bibitem [{\citenamefont {Boyd}(1995)}]{Boyd95}%
  \BibitemOpen
  \bibfield  {author} {\bibinfo {author} {\bibfnamefont {J.~P.}\ \bibnamefont
  {Boyd}},\ }\bibfield  {title} {\bibinfo {title} {Multiple precision
  pseudospectral computations of the radiation coefficient for weakly nonlocal
  solitary waves: Fifth-order {Korteweg--DeVries} equation},\ }\href
  {https://doi.org/10.1063/1.168557} {\bibfield  {journal} {\bibinfo  {journal}
  {Computers in Physics}\ }\textbf {\bibinfo {volume} {9}},\ \bibinfo {pages}
  {324} (\bibinfo {year} {1995})}\BibitemShut {NoStop}%
\bibitem [{\citenamefont {Boyd}(2013)}]{Boyd13}%
  \BibitemOpen
  \bibfield  {author} {\bibinfo {author} {\bibfnamefont {J.}~\bibnamefont
  {Boyd}},\ }\href {https://store.doverpublications.com/0486411834.html} {\emph
  {\bibinfo {title} {Chebyshev and Fourier Spectral Methods: Second Revised
  Edition}}},\ Dover Books on Mathematics\ (\bibinfo  {publisher} {Dover
  Publications},\ \bibinfo {year} {2013})\BibitemShut {NoStop}%
\bibitem [{cln()}]{clnweb}%
  \BibitemOpen
  \href@noop {} {\bibinfo {title} {{CLN} - {Class Library for Numbers}}},\
  \bibinfo {howpublished} {\url{https://www.ginac.de/CLN/}}\BibitemShut
  {NoStop}%
\bibitem [{arb()}]{arbweb}%
  \BibitemOpen
  \href@noop {} {\bibinfo {title} {{Arb} - {a C library for arbitrary-precision
  ball arithmetic}}},\ \bibinfo {howpublished}
  {\url{https://arblib.org/}}\BibitemShut {NoStop}%
\bibitem [{\citenamefont {Johansson}(2017)}]{Johansson17}%
  \BibitemOpen
  \bibfield  {author} {\bibinfo {author} {\bibfnamefont {F.}~\bibnamefont
  {Johansson}},\ }\bibfield  {title} {\bibinfo {title} {Arb: efficient
  arbitrary-precision midpoint-radius interval arithmetic},\ }\href
  {https://doi.org/10.1109/TC.2017.2690633} {\bibfield  {journal} {\bibinfo
  {journal} {IEEE Transactions on Computers}\ }\textbf {\bibinfo {volume}
  {66}},\ \bibinfo {pages} {1281} (\bibinfo {year} {2017})}\BibitemShut
  {NoStop}%
\bibitem [{\citenamefont {Press}\ \emph {et~al.}(2007)\citenamefont {Press},
  \citenamefont {Teukolsky}, \citenamefont {Vetterling},\ and\ \citenamefont
  {Flannery}}]{numrec07}%
  \BibitemOpen
  \bibfield  {author} {\bibinfo {author} {\bibfnamefont {W.}~\bibnamefont
  {Press}}, \bibinfo {author} {\bibfnamefont {S.}~\bibnamefont {Teukolsky}},
  \bibinfo {author} {\bibfnamefont {W.}~\bibnamefont {Vetterling}},\ and\
  \bibinfo {author} {\bibfnamefont {B.}~\bibnamefont {Flannery}},\ }\href
  {http://numerical.recipes/} {\emph {\bibinfo {title} {Numerical Recipes 3rd
  Edition: The Art of Scientific Computing}}}\ (\bibinfo  {publisher}
  {Cambridge University Press},\ \bibinfo {year} {2007})\BibitemShut {NoStop}%
\bibitem [{\citenamefont {Sun}(1998)}]{Sun98}%
  \BibitemOpen
  \bibfield  {author} {\bibinfo {author} {\bibfnamefont {S.~M.}\ \bibnamefont
  {Sun}},\ }\bibfield  {title} {\bibinfo {title} {On the oscillatory tails with
  arbitrary phase shift for solutions of the perturbed {Korteweg--de Vries}
  equation},\ }\href {https://doi.org/10.1137/S0036139996299212} {\bibfield
  {journal} {\bibinfo  {journal} {SIAM Journal on Applied Mathematics}\
  }\textbf {\bibinfo {volume} {58}},\ \bibinfo {pages} {1163} (\bibinfo {year}
  {1998})}\BibitemShut {NoStop}%
\end{thebibliography}%

\end{document}